\documentclass[aps,superscriptaddress,twocolumn,twoside,floatfix,prx,nofootinbib,a4paper]{revtex4-2}
\usepackage{times}
\usepackage{amsmath,amssymb,amsthm,mathtools,thmtools,nameref}
\usepackage{comment}
\usepackage{enumerate}
\usepackage{float}
\usepackage{xcolor}
\usepackage[colorlinks=true,linkcolor=blue,citecolor=magenta,urlcolor=blue]{hyperref}
\usepackage[framemethod=default]{mdframed}
\usepackage{csvsimple}
\usepackage[mode=tex]{standalone}
\usepackage[capitalize]{cleveref}

\usepackage{pgfplots}
\pgfplotsset{compat=1.18} 
\usepackage{tikz}
\usepackage{quantikz}

\newcommand{\tr}{\text{Tr}}

\global\long\def\isdef{\coloneqq}%
\global\long\def\kb#1#2{\left|#1\vphantom{#2}\right\rangle \left\langle \vphantom{#1}#2\right|}%
\global\long\def\braket#1#2{\left\langle #1\middle|#2\right\rangle }%
\global\long\def\braOket#1#2#3{\left\langle #1\middle|#2\middle|#3\right\rangle }%
\global\long\def\Tr{\operatorname{Tr}}%

\newtheorem{theorem}{Theorem}

\newtheorem{lemma}[theorem]{Lemma}

\begin{document}
\title{Efficient Long-Range Entanglement using Dynamic Circuits}
\author{Elisa~B\"aumer}
\email{eba@zurich.ibm.com}
\affiliation{IBM Quantum, IBM Research -- Zurich, 8803 R\"uschlikon, Switzerland}
\author{Vinay~Tripathi}
\affiliation{IBM Quantum, IBM T.J. Watson Research Center, Yorktown Heights, NY 10598, USA}
\affiliation{Department of Physics \& Astronomy, University of Southern California,
Los Angeles, California 90089, USA}
\author{Derek~S.~Wang}
\affiliation{IBM Quantum, IBM T.J. Watson Research Center, Yorktown Heights, NY 10598, USA}
\author{Patrick~Rall}
\affiliation{IBM Quantum, MIT-IBM Watson AI Lab, Cambridge, MA 02142, USA}
\author{Edward~H.~Chen}
\affiliation{IBM Quantum, Almaden Research Center, San Jose, CA 95120, USA}
\affiliation{IBM Quantum, Research Triangle Park, NC 27709, USA}
\author{Swarnadeep~Majumder}
\affiliation{IBM Quantum, IBM T.J. Watson Research Center, Yorktown Heights, NY 10598, USA}
\author{Alireza Seif}
\affiliation{IBM Quantum, IBM T.J. Watson Research Center, Yorktown Heights, NY 10598, USA}
\author{Zlatko~K.~Minev}
\affiliation{IBM Quantum, IBM T.J. Watson Research Center, Yorktown Heights, NY 10598, USA}

\begin{abstract}
Quantum simulation traditionally relies on unitary dynamics, inherently imposing efficiency constraints on the generation of intricate entangled states. In principle, these limitations can be superseded by non-unitary, dynamic circuits. These circuits exploit measurements alongside conditional feed-forward operations, providing a promising approach for long-range entangling gates, higher effective connectivity of near-term hardware, and more efficient state preparations.
Here, we explore the utility of shallow dynamic circuits for creating long-range entanglement on large-scale quantum devices. Specifically, we study two tasks: \mbox{CNOT} gate teleportation between up to 101 qubits by feeding forward 99 mid-circuit measurement outcomes, and the preparation of Greenberger–Horne–Zeilinger (GHZ) states with genuine entanglement.
In the former, we observe that dynamic circuits can outperform their unitary counterparts. In the latter, by tallying instructions of compiled quantum circuits, we provide an error budget detailing the obstacles that must be addressed to unlock the full potential of dynamic circuits. Looking forward, we expect dynamic circuits to be useful for generating long-range entanglement in the near term on large-scale quantum devices.

\vspace{1cm}
\end{abstract}

\maketitle

\section{Introduction}
Quantum systems present two distinct modes of evolution: deterministic unitary evolution, and stochastic evolution as the consequence of quantum measurements. To date, quantum computations predominantly utilize unitary evolution to generate complex quantum states for information processing and simulation. However, due to inevitable errors in current quantum devices~\cite{Preskill2018}, the computational reach of this approach is constrained by the depth of the quantum circuits that can realistically be implemented on noisy devices. The introduction of non-unitary dynamic circuits, also called adaptive circuits or LAQCC (local alternating quantum classical computation) circuits \cite{buhrman2023state}, can not only implement more general quantum channels, but may also be able to overcome some of these limitations by employing mid-circuit measurements and feed-forward operations. As classical computation and communication are viewed as essentially free compared to quantum operations, such conditional operations are a necessary ingredient for quantum error correction (see, \textit{e.g.}, Ref.~\cite{Terhal2015}).
In the near term, dynamic circuits present a promising avenue for generating long-range entanglement, a task at the heart of quantum algorithms \cite{hoyer2003, pham2013}. This includes both implementation of long-range entangling gates that, due to local connectivity among the qubits in many quantum platforms, can require deep unitary quantum circuits, and preparation of many-qubit entangled~\cite{GHZ1989,raussendorf2005longrange} and topologically ordered quantum states~\cite{Dennis2002,piroli2021locc,kim2021measurement,verresen2112efficiently,tantivasadakarn2021long,tantivasadakarn2022shortest,tantivasadakarn2023hierarchy,bravyi2022adaptive,PRXQuantum.4.030318}.

From a physical standpoint, the entanglement needs to propagate across the entire range between the qubits. Given that the entanglement cannot spread faster than its information light cone~\cite{Lieb1972, Brayvi2006}, entangling two qubits that are a distance $n$ apart requires a minimum two-qubit gate-depth that scales as $\mathcal{O}(n)$, and even when assuming all-to-all connectivity, the generation of entanglement over $n$ qubits necessitates a minimum two-qubit gate depth of $\mathcal{O}(\log n)$. Thus, the task becomes challenging when applying only unitary gates. Using dynamic circuits, the spread of information can be mostly conducted by classical calculations, which can be faster and with a higher fidelity than the quantum gates, and long-range entanglement can be created in a shallow quantum circuit \cite{Raussendorf2003,Jozsa2005,Beverland_2022}, \textit{i.e.} the depth of quantum gates is constant for any $n$. 

While dynamic circuits have been explored in small-scale experiments~\cite{Barrett2004, Pfaff2012, Rist2013, Wan2019, Corcoles2021}, only recently have there been experimental capabilities on large-scale quantum devices. However, most demonstrations (with the exception of \textit{e.g.} Refs.~\cite{Fossfeig2023,Iqbal2023,moses2023race,baeumer2024quantum}) have utilized post-selection~\cite{Cao2023} or post-processing~\cite{PRXQuantum.4.020315,chen2023nishi} instead of feed-forward to prepare entangled states. Such approaches enable the study of properties of the state prepared in isolation, but have limited applicability when the state preparation is part of a larger quantum information processing task. 

Here, we explore the utility of shallow dynamic circuits for creating long-range entanglement on large-scale superconducting quantum devices. 
In Section~\ref{sec:entgates}, we demonstrate an advantage with dynamic circuits by teleporting a long-range entangling \mbox{CNOT} gate over up to 101 locally connected superconducting qubits. We also discuss how this approach can be generalized to more complex gates, such as the three-qubit \mbox{Toffoli} gate. Then, in Section~\ref{sec:stateprep}, we prepare a long-range entangled state, the GHZ state \cite{GHZ1989}, with a dynamic circuit. We show that---with a composite error mitigation stack customized for the hardware implementation of dynamic circuits---we can prepare genuinely entangled GHZ states but fall short of state-of-the-art system sizes achieved with unitary gates due to hardware limitations. We predict conditions under which dynamic circuits should be advantageous over unitary circuits based on our error budget calculation.

\begin{figure*}[htb]
\includegraphics[width=1.95\columnwidth]
 {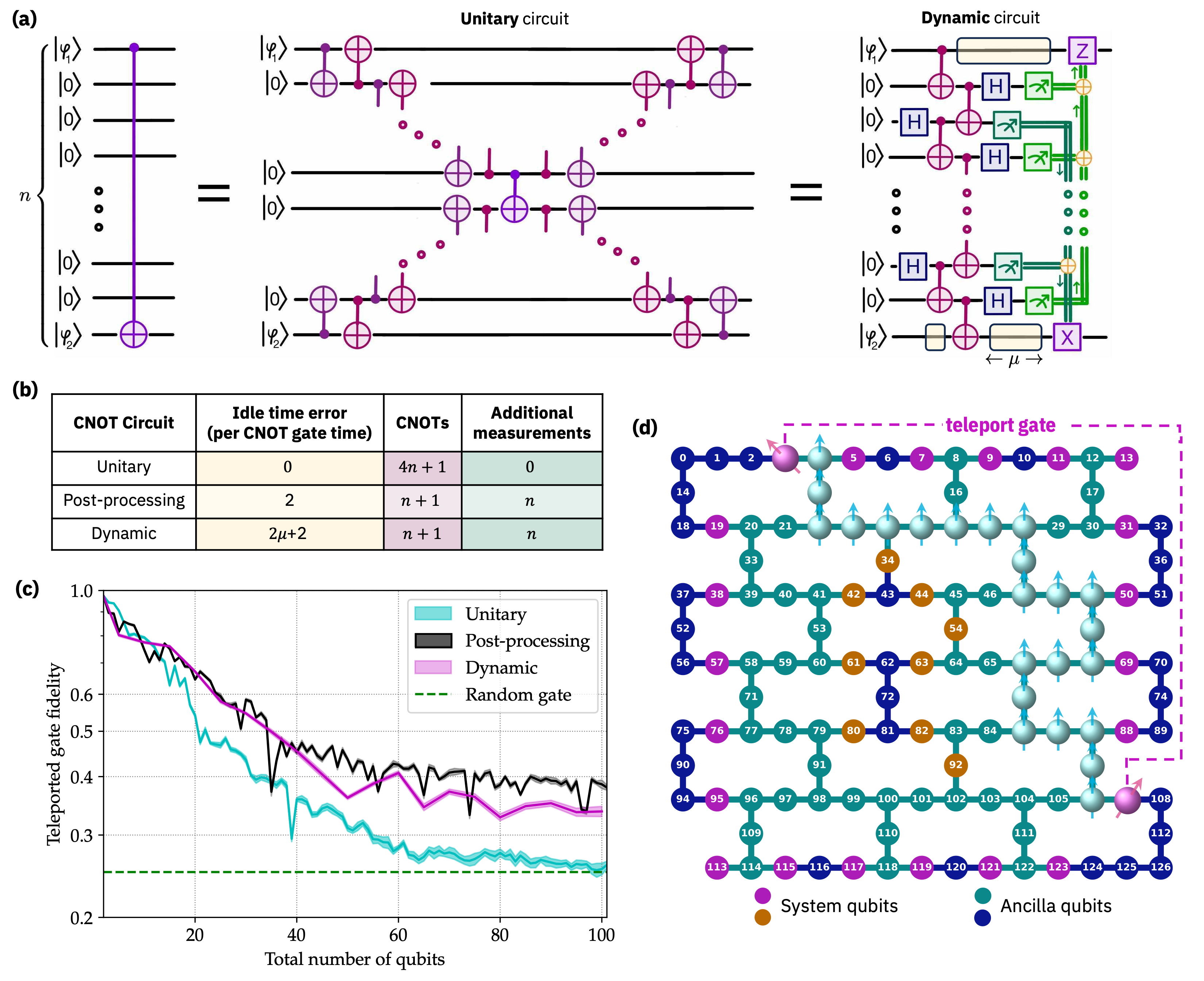}
\caption{
Teleporting a~\mbox{CNOT} gate for long-range entanglement. 
(a) 
Left: Circuit for a long-range \mbox{CNOT} gate spanning a 1D chain of $n$-qubits subject to nearest-neighbor connections only. 
Middle: Equivalent unitary decomposition into implementable \mbox{CNOT} gates; circuit  depth~$\mathcal{O}(n)$. 
Right: Equivalent circuit employing measurements with feed-forward operations; circuit depth~$\mathcal{O}(1)$.
If the post-measurement state is unused, feed-forward operations can be handled in post-processing, eliminating the need for their experimental implementation.
Yellow regions indicate the idle time during CNOT gates on other qubits as well as during measurement and feed-forward (which is denoted by duration~$\mu$).
(b) 
Error model inputs for unitary, measurement-based, and dynamic-circuit CNOT protocols comprise the total number of: non-zero idle-block times, CNOT gates, and additional measurements.
(c) Experimental results, where dynamic circuits offer improved fidelity for \mbox{CNOT} gate teleportation across a qubit chain $\gtrsim$ 10 qubits.
(d) Map of a 127-qubit heavy-hexagonal processor, \texttt{ibm\_sherbrooke}, overlaid with system configurations for long-range gate teleportation across a locally connected bus. 
To establish an effective all-to-all connectivity, we show one possible strategy of dividing the qubits into system (purple and orange) and sacrificial ancilla (turquoise and blue for extra connections) qubits. 
To parallellize gate execution with increased connectivity, orange qubits can be used as ancillas. We show how a particular long-range \mbox{CNOT} can be implemented through an ancilla bus marked as turquoise spins.
} 
\label{fig:cnot}
\end{figure*}

\section{CNOT gate teleportation}
\label{sec:entgates}
The limited connectivity between qubits in many quantum computational platforms can result in the compilation of non-local unitary circuits into deep and error-prone unitary circuits. A potential solution is the use of shallow dynamic circuits. The crucial ingredient for such protocols is long-range \mbox{CNOT} gates from the first to $n$th qubit, as shown on the left in Fig.~\ref{fig:cnot}(a). In the following, we demonstrate a regime under which dynamic circuits enable higher-fidelity long-range \mbox{CNOT} gates \textit{via} gate teleportation. We first describe the dynamic circuit and compare to its equivalent unitary counterpart. We argue, using a simple error budget, that there exists a regime in which the dynamic circuit implementation has an advantage over the unitary one, see Fig.~\ref{fig:cnot}(b). Then, using up to $101$ qubits on a superconducting processor, we demonstrate a crossover in the fidelity of \mbox{CNOT} gate teleportation, where dynamic circuits perform better for entangling qubits over longer ranges; see Fig.~\ref{fig:cnot}(c). 
This gate teleportation scheme enables an effective all-to-all connectivity in devices with a more limited connectivity, such as those on a heavy-hexagonal lattice. By using some of the qubits as ancillas for measurement and classical feed-forward operations, the ancilla qubits form a bus that connects all system qubits with each other. Therefore, by sacrificing some of the qubits in a large device with limited connectivity, we gain effective access to an all-to-all connected device with fewer qubits; see Fig~\ref{fig:cnot}(d). As this effective all-to-all connectivity limits the parallelization of gates, the orange system qubits could be sacrificed as ancilla qubits as well to further parallelize gate execution with increased connectivity. In addition, a clever compilation could increase parallelization, as e.g. shown in Appendix~\ref{app:toffoli} in Fig.~\ref{fig:ccz_circuit}, where a long-range CCZ gate could be implemented with two feed-forward operations rather than teleporting all six CNOT gates separately. 

We describe the dynamic circuit for \mbox{CNOT} gate teleportation, shown on the right in Fig.~\ref{fig:cnot}(a) and derived in Appendix \ref{app:cnotderiv}.
Importantly, this dynamic circuit can be straightforwardly extended for any number of qubits $n$ (where $n$ is the number of ancillas) such that the depth remains constant for any initial states $\ket{\varphi_1}$ $\left(\ket{\varphi_2}\right)$ of the control (target) qubit. We expect the error to be dominated by the $n$ mid-circuit measurements, $n+1$ \mbox{CNOT} gates parallelized over 2 gate layers, and idle time mostly over the classical feed-forward time.
Note that in this particular realization, each of the $n$ ancilla qubits between the two system qubits must be in the state $\ket{0}$. Therefore, during the course of the gate teleportation, the ancillas cannot also be used as memory qubits, further motivating the division of qubits into system and sacrificial ancilla qubits in Fig.~\ref{fig:cnot}(d).  

We also present an equivalent, low-error unitary counterpart in the middle of Fig.~\ref{fig:cnot}(a). (In Appendix~\ref{app:comparisoncnot}, we propose several different unitary implementations of the long-range \mbox{CNOT} gate. Based on experimental results, as well as the noise model described in  Appendix~\ref{app:noisemodel} that gives rise to the error budget described in Appendix~\ref{app:cnot_errorbudget}, we select this one.) In this unitary realization, the system qubits are connected by a bus of ancilla qubits that are initialized in and returned to the $|0\rangle$ state, just as in its dynamic counterpart. In our particular compilation, throughout the execution of the circuit, qubits that are not in the $|\phi_1\rangle$ or $|\phi_2\rangle$ state are in the $|0\rangle$ state. Doing so minimizes both decoherence and cross-talk errors intrinsic to our superconducting qubit design, as heuristically we learned that the noise affecting our qubits is primarily limited to amplitude damping, dephasing and ZZ crosstalk errors on neighboring qubits, which implies essentially no idling errors on qubits in the $\ket{0}$ state. Therefore, relative to the dynamic version, there is no error due to idle time or mid-circuit measurements, although there are $\sim$4 times more \mbox{CNOT} gates.

A summary of the error budgets for the dynamic and unitary circuits is in Fig.~\ref{fig:cnot}(b). Based on this table, we expect that dynamic circuits should be advantageous over unitary circuits if the additional $n$ mid-circuit measurements in the dynamic circuit introduce less error than the $3n$ extra \mbox{CNOT} gates in the unitary circuit, assuming $n$ is large enough such that the idling error $\mu$ incurred during measurement and classical feed-forward in the dynamic circuit is relatively small.
Importantly, we should note that these error analyses only consider the gate error on the two respective qubits, but not the error introduced on other qubits, which we expect to be much larger in the unitary case due to the linear depth. Thus, the constant-depth dynamic circuit might be even more advantageous than what we can see from the gate fidelity.

To determine the experimental gate fidelity, let our ideal unitary channel be~$\mathcal{U}(\rho)\isdef U\rho U^{\dagger}$ and its noisy version be $\tilde{\mathcal{U}}(\rho)\isdef\mathcal{U}(\Lambda(\rho))$,
where $\Lambda$ is the effective gate noise channel and $\rho$ is a quantum state. The average gate fidelity of the noisy gate is~$\mathcal{F}_{\mathrm{avg}}\left(\mathcal{U},\tilde{\mathcal{U}}\right) = \int\mathrm{d}\psi\,\Tr\left[\mathcal{U}\left(\rho_\psi\right)\tilde{\mathcal{U}}\left(\rho_\psi\right)\right]$, where the Haar average is taken over the pure states~$\rho_\psi = \kb{\psi}{\psi}$. This fidelity can be faithfully estimated from Pauli measurements on the system, using \textit{Monte Carlo process certification} \cite{Flammia2011,daSilva2011}, as detailed in  Appendix~\ref{app:gatefidelity}.

The results from a superconducting quantum processor are shown in Fig.~\ref{fig:cnot}(c). The implementation details can be found in Appendix~\ref{app:expdetails_cnot}. As expected, for a small number of qubits $n\lesssim 10$ the unitary implementation yields the best fidelities. However, for increasing $n$ it converges much faster to the fidelity of a random gate (0.25) than the dynamic circuits implementation, which converges to a value slightly below 0.4. These align well with the error budget analysis in  Appendix~\ref{app:cnot_errorbudget} and the noise model predictions depicted in  Appendix~\ref{app:noisemodel}. Note that, in the limit of large $n$, the fidelities of the measurement-based scheme are limited by the $Z$ and $X$ corrections on $\ket{\phi_1}$ and  $\ket{\phi_2}$ (see Fig.~\ref{fig:cnot}(a)). A straightforward derivation using this noise model shows that the minimum possible process fidelity due to only incorrect $Z$ and $X$ corrections (without the fixed infidelity from the idle time and CNOT gates) is 0.25, which converts to a gate fidelity of 0.4.

The measurement-based protocol with post-processing performs slightly better than the dynamic circuits as the former does not incur errors from the classical feed-forward, allowing us to isolate the impact of classical feed-forward from other errors, such as the $n+1$ intermediate \mbox{CNOT} gates and mid-circuit measurements. Note, however, that the post-processing approach is generally not scalable if further circuit operations follow the teleported \mbox{CNOT} due to the need to simulate large system sizes, further emphasizing the advantage of dynamic circuits as errors rooted in classical feedforward are reduced. Overall, we find that \mbox{CNOT} gates over large distances are more efficiently executed with dynamic circuits than unitary ones.

In Appendix~\ref{app:toffoli} we show that these ideas can be generalized to teleporting multi-qubit gates, such as the \mbox{Toffoli} or \mbox{CCZ} gate. Compiling them more efficiently than simply implementing multiple teleported \mbox{CNOT} gates, we expect their shallow implementation with dynamic circuits to be even more advantageous over their unitary counterpart, especially for large $n$.

\section{State preparation: GHZ} \label{sec:stateprep}

\begin{figure*}[htb]
	\centering
 \includegraphics[width=1.95\columnwidth]{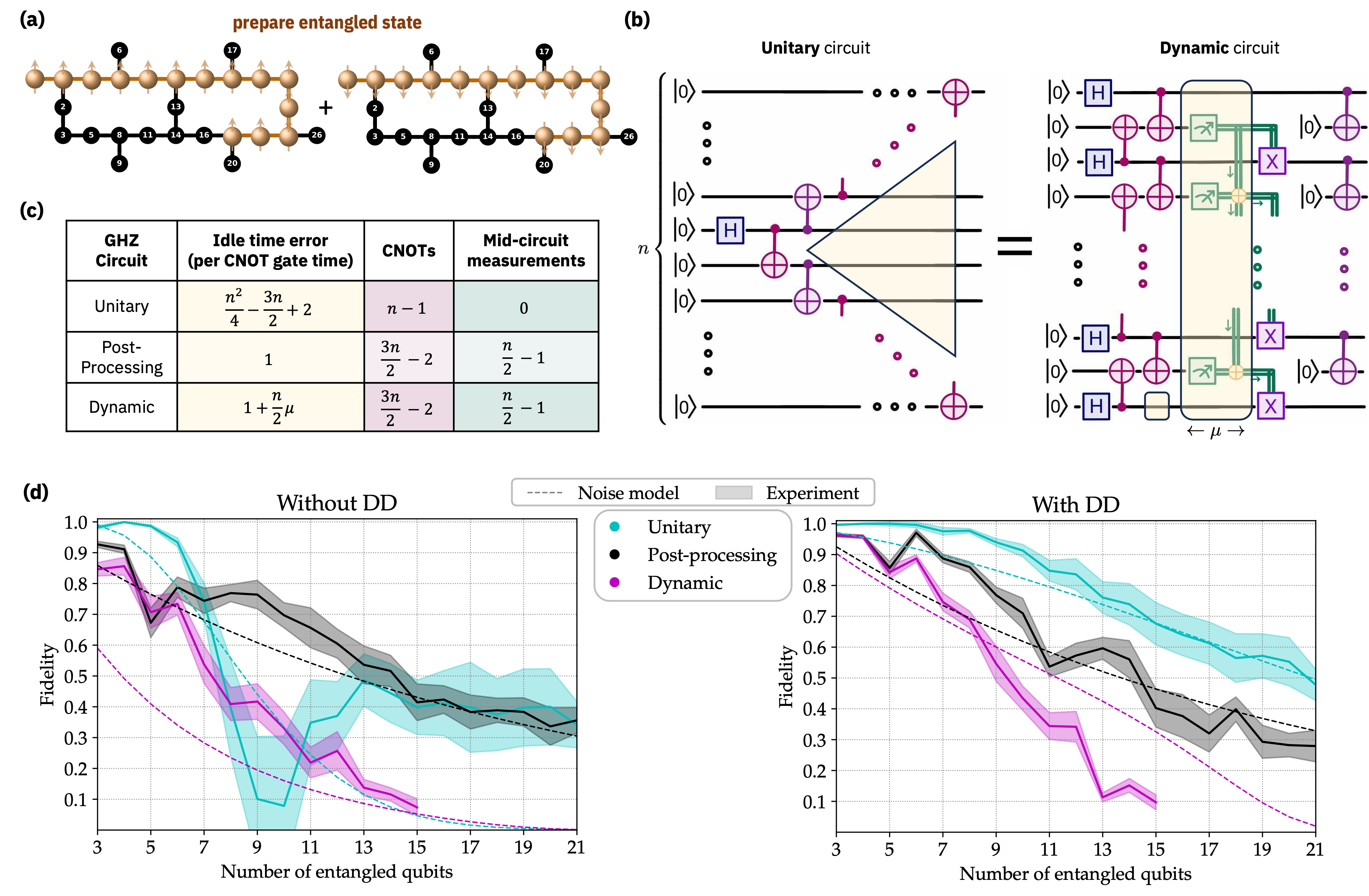}
\caption{
Preparing long-range entangled states. 
(a) Illustration of a GHZ state with chosen qubit spins (spheres) in a superposition of ``all up" and ``all down" polarizations (arrows), overlaid on a quantum processor.
(b) Circuits to prepare an $n$-qubit GHZ state using either a unitary (left) or dynamic (right) circuit. For a 1D qubit chain, the depth of the unitary (resp., dynamic) circuit scales as $\mathcal{O}(n)$ (resp., $\mathcal{O}(1)$). 
If the final state is not directly used, the feed-forward operations can be implemented in classical post-processing on the output bits (classically controlled X gates and resets can be omitted). Yellow regions indicate the idle time during CNOT gates on other qubits as well as during measurement and feed-forward (which is denoted by duration~$\mu$).
(c) 
Error model inputs for the GHZ preparation circuits. 
The model incorporates the noisy components of the circuits: non-zero idle circuit periods (yellow), number of \mbox{CNOT} gates (pink), and the number of mid-circuit measurements (green). 
These parameters are used to derive an error model that yields a lower-bound on the protocol fidelity, shown in the following panel.
(d) 
Fidelity of preparing the GHZ state on quantum hardware using unitary, measurement-based post-processing, or dynamic circuits in the absence or presence of dynamical decoupling (DD). Data shown with dots. Theory curves based on the error model parameters of panel (c) shown in dashed lines.
 }
	\label{fig:ghz}
\end{figure*}

Dynamic circuits can also be used to prepare long-range entangled states. A prototypical example is the GHZ state \cite{GHZ1989}, shown schematically in Fig.~\ref{fig:ghz}(a). While it can be created using only Clifford gates and thus can be simulated efficiently on a classical computer \cite{Gottesman1998}, it becomes non-simulatable when followed by a sufficient number of non-Clifford gates in a larger algorithm, or when inserted as a crucial ingredient in \textit{e.g.} efficient compilation of multi-qubit gates \cite{Hoyer_2005, yang2023harnessing}.

Here, we show that GHZ states with long-range entanglement can be prepared with dynamic circuits. Although we do not see a clear advantage of dynamic circuits over unitary ones in this case, we provide a detailed description of the challenges that must be addressed to realize such an advantage.

For preparation of a GHZ state on a 1D $n$-qubit chain, in Fig.~\ref{fig:ghz}, we show the equivalence between the unitary circuit (left) and dynamic circuit (right). (For a detailed derivation, see Appendix \ref{app:ghzderiv}.)
Notably, the unitary equivalent has a two-qubit gate depth that scales as $\mathcal{O} \left(n\right)$ with quadratically increasing idle time and $n-1$ total \mbox{CNOT} gates, while the depth of the dynamic circuits remains constant with linearly increasing idle time, $3n/2-1$ total \mbox{CNOT} gates, and $n/2-1$ mid-circuit measurements (see Fig.~\ref{fig:ghz}(c)). The dynamic circuit incurs less idle time and fewer two-qubit gate depth at the cost of increased \mbox{CNOT} gates and mid-circuit measurements. Therefore, we expect dynamic circuits to be advantageous for large system sizes $n$ and low errors in mid-circuit measurement. For a more detailed analysis of the error budget, see Appendix~\ref{app:GHZerrorbudget}.

We explore whether current large-scale superconducting quantum devices enable an advantage with dynamic circuits for preparation of the entangled GHZ state. 
To efficiently verify the preparation of a quantum state $\sigma$, we use the Monte Carlo state certification that samples from Pauli operators with non-zero expectation values, as implemented in Ref.~\cite{Cao2023} and described in detail in  Appendix~\ref{app:statefidelity}.
As the $n$-qubit GHZ state is a stabilizer state, we can randomly sample $m$ of the $2^n$ stabilizers $\{S_i\}_{i = 1..2^n}$ and approximate the fidelity by $F = \frac{1}{m}\sum_{k=1}^m \langle S_k \rangle_\sigma + \mathcal{O}\left(\frac{1}{\sqrt{m}}\right)$.

The experimental results of GHZ state preparation with unitary and dynamic circuits are shown in Fig.~\ref{fig:ghz}(d). They all include measurement error mitigation on the final measurements~\cite{Nation2023Mapomatic}. The implementation details can be found in Appendix~\ref{app:expdetails_ghz}. On the left, we show the results without dynamical decoupling. 
In the unitary case, we observe genuine multipartite entanglement, defined as state fidelity $F>0.5$~\cite{Leibfried2005},  within a confidence interval of $95\%$ up to 7 qubits with a rapid decay in fidelity with increasing system size mainly due to errors in two-qubit gates and ZZ crosstalk errors during idling time \cite{Tripathi2022}. As these errors are mostly coherent, they lead to an oscillation of the fidelity such that it increases again for higher qubit numbers. To suppress the coherent ZZ errors we apply dynamical decoupling (DD) pulses, as described below. 

In the dynamic case, we observe genuine entanglement up to 6 qubits. Here, we do not find a crossover point after which dynamic circuits have an advantage over unitary circuits. We attribute the performance of dynamic circuits to several factors, including the fact that the current implementation results in an average classical feedforward time that scales with the number of potential mid-circuit measurement bitstring outcomes, which itself grows exponentially with system size. This limitation appears because the switch operator is currently testing each possible case of measurement outcomes sequentially, so on average checks half of the cases until it finds the correct one. With our future control software we expect to implement the correct feed-forward operations in constant time.
By reducing the error induced by idle time during classical feedforward, we expect dynamic circuits to surpass unitary circuits at $\gtrsim$10 qubits---we can see this by studying the post-processing case, which is equivalent to the dynamic circuit implementation except that the classical logic is executed in post-processing, not during execution of the quantum circuit itself.

On the right of Fig.~\ref{fig:ghz}(d), we show the results using dynamical decoupling~(DD)~\cite{Viola1999DD, Jurcevic2021}. We observe improved fidelities for both the unitary and dynamic circuit cases, but not for the post-processing case as there is little error induced by idle times to quench with dynamical decoupling in the first place. For the unitary case, we observe genuine multipartite entanglement up to 17 qubits, more than twice as many compared to the unmitigated unitary case. This result is close to the state of the art on superconducting quantum processors and is limited by the fact that we do not leverage the 2D connectivity of the device, as in Ref.~\cite{Mooney2021}. While the fidelities are improved with DD for dynamic circuits, the improvement is less dramatic. We attribute this difference to two reasons: First, the unitary circuit has a quadratic idling error term in contrast to a leading linear term for dynamic circuits, resulting in comparatively smaller improvement for dynamic circuits with dynamical decoupling. Second, with the current controls, we are not able to apply DD pulses during the classical feedforward time, which is the main source of idling error in the dynamic circuit. As in the unmitigated case, we observe rapid decay of the fidelity with increasing system. This can again be partially attributed to exponential growth of the classical feedforward time.
In the future, we expect to reduce this scaling to constant, in which case we expect drastically improved performance and genuine entanglement up to $\sim$15 qubits. Still, however,  we do not expect to observe an advantage with dynamic circuits for preparation of GHZ states over unitary ones. To realize an advantage with dynamic circuits, we require a scenario where the quadratically scaling idle error of the unitary circuit dominates over sufficiently small CNOT and mid-circuit measurement error; see  Appendix~\ref{app:crossover} for a more detailed analysis. We anticipate these conditions can be realized through a combination of hardware improvements and the extension of error mitigation techniques, such as probabilistic error cancellation~\cite{Temme2017PECandZNE, Berg2022}, toward mid-circuit measurements.

\section{Conclusion and Outlook} \label{sec:conclusion}
Dynamic circuits are a promising feature toward overcoming connectivity limitations of large-scale noisy quantum hardware. Here, we demonstate their potential for efficiently generating long-range entanglement with two useful tasks: teleporting entangling gates over long ranges to enable effective all-to-all connectivity, and state preparation with the GHZ state as an example. For \mbox{CNOT} gate teleportation, we show a regime in which dynamic circuits result in higher fidelities on up to 101 qubits of a large-scale superconducting quantum processor. 
We leave incorporating this more efficient implementation of long-range entangling gates as a subroutine in another quantum algorithm to future work; potential studies can include simulating many-body systems with non-local interactions. As we demonstrate theoretically, gate teleportation schemes can be extended beyond \mbox{CNOT} gates to multi-qubit ones, such as the \mbox{CCZ} gate. Its experimental implementation is also a promising project for the future.
For state preparation, based on both unmitigated and mitigated hardware experiments, we expect to see the value of dynamic circuits once the classical post-processing becomes more efficient and the mid-circuit measurement errors can be reduced. We plan on revising the experiments as soon these capabilities are available. We anticipate that further experiments with dynamic circuits and the development of noise models describing them will be vital contributions toward efficient circuit compilation, measurement-based quantum computation, and fault-tolerant quantum computation.

\section{Acknowledgements}

We thank Diego Rist\`{e}, Daniel Egger, Alexander Ivrii and Luke Govia for valuable discussions and feedback. We thank Emily Pritchett, Maika Takita, Abhinav Kandala, and Sarah Sheldon for their help. We also thank Thomas Alexander, Marius Hillenbrand, and Reza Jokar for their support with implementing dynamic circuits.

\appendix
\onecolumngrid
\newpage

\section{Circuit Derivations}

\begin{figure}[htb]
	\centering
	\includegraphics[width=0.65\columnwidth]{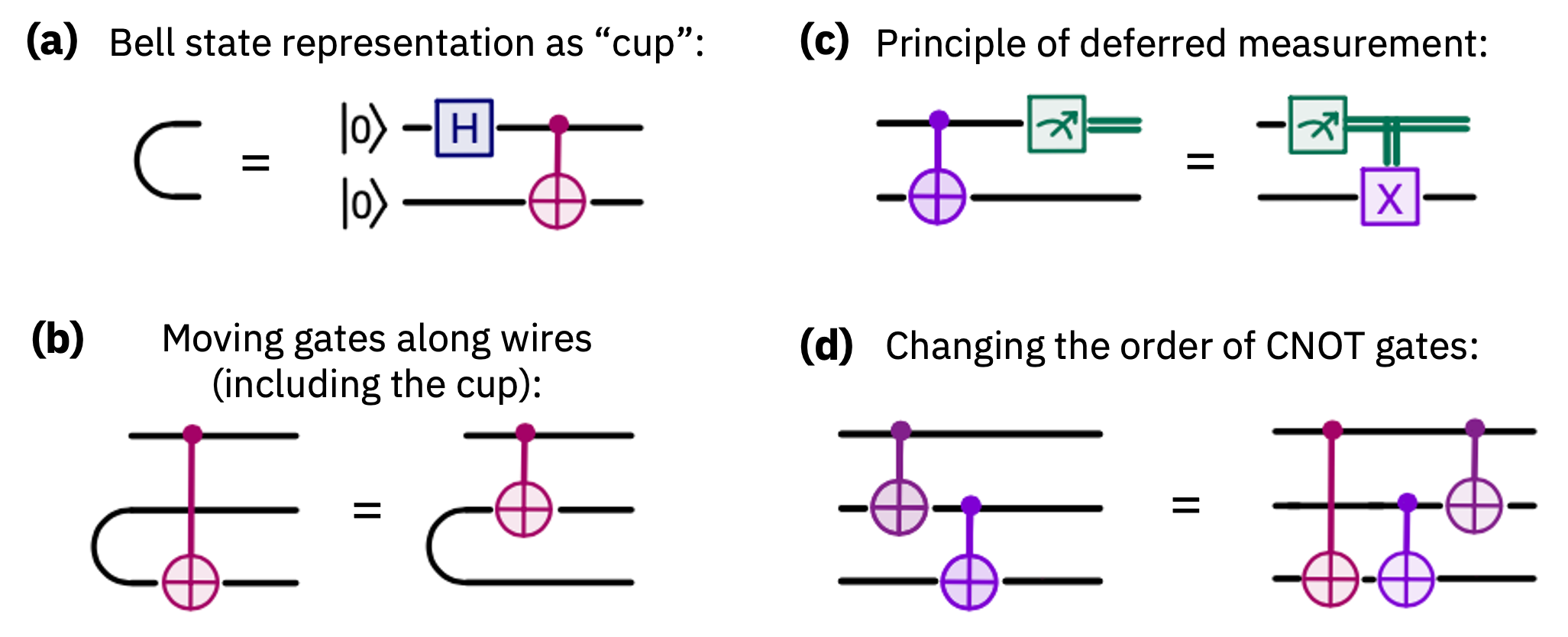}
	\caption{Useful circuit identities that are used in the illustrative derivation of the CNOT gate teleportation and GHZ state preparation.}
	\label{fig:circuit_identities}
\end{figure}

In the following we show the circuit equivalences of the CNOT gate teleportation (Fig.~\ref{fig:cnot}(a)) and the GHZ state preparation (Fig.~\ref{fig:ghz}(b)). We are not claiming any novelty with this ``proof", but just wanted to show the reader how to derive them in an illustrative way. Before, let us start with some features that we will be using:

\begin{itemize}
\item The Bell state $\frac{1}{\sqrt{2}}(\ket{00}+\ket{11})$ can be illustrated as a so-called ``cup", as shown in Fig.~\ref{fig:circuit_identities}(a), We can move gates along wires including along the cup, as in Fig.~\ref{fig:circuit_identities}(b).
\item \textit{Principle of deferred measurement}: a controlled gate followed by a measurement of the controlled qubit results in the same as first performing the measurement and then applying a classically-controlled gate as in Fig.~\ref{fig:circuit_identities}(c).
\item While CNOT gates commute when they are conditioned on the same qubit or have the same target qubit, we get an extra gate when they act on the same qubit differently as shown in Fig.~\ref{fig:circuit_identities}(d).
\end{itemize}

\subsection{Long-Range CNOT} \label{app:cnotderiv}

\begin{figure}[htb]
	\centering
	\includegraphics[width=0.95\columnwidth]{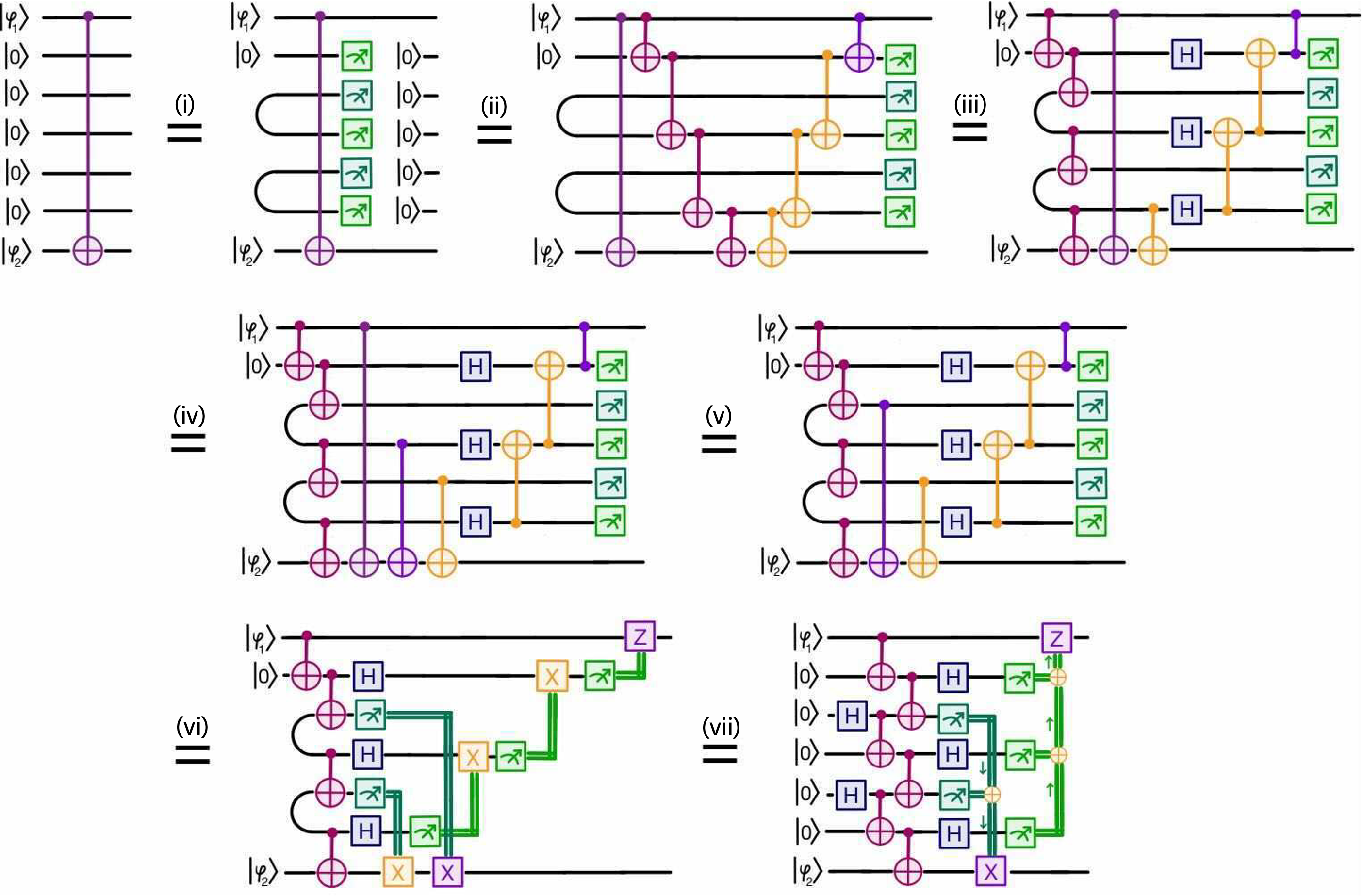}
	\caption{Graphical derivation for reducing a long-range CNOT gate into gate teleportation executed with measurements and feed-forward operations, \textit{i.e.}, a dynamic circuit. Roman numerals indicate sequential step numbers described in main text.}
	\label{fig:cnot_derivation}
\end{figure}

In Fig.~\ref{fig:cnot_derivation} we illustrate a derivation of the CNOT gate teleportation, as exemplified for $7$ qubits, which can be straightforwardly extended to an arbitrary number of qubits. In the following, we provide explanations for each step of the derivation, labeled by roman numerals in the figure:
\begin{enumerate}[(i)]
\item In the first step, we observe that entangling, measuring, and resetting the ancilla qubits does not affect the circuit.
\item We insert CNOT gates that would cancel each other. From now on we omit to write down the reset of the ancilla qubits following the measurement.
\item We move the pink CNOT gates along the Bell states to the respective qubits above. Also, we add Hadamard gates to flip the direction of the orange CNOT gates (except for the one at the bottom). Note that we can omit the hadamard gates right before the measurements, as they are not affecting the other qubits anymore.
\item By moving the bottom orange CNOT ``up" along the Bell state and passing a pink CNOT, we get the extra purple CNOT gate.
\item Moving the new purple CNOT ``up" along the Bell state, an extra gate appears that cancels with the initial long-range CNOT gate when pushed to the left (and then it is controlled on state $\ket{0}$, so can be omitted as well).
\item Now we make use of the principle of deferred measurement.
\item In a final step we merge the classically-conditioned gates. 
The orange $\oplus$ correspond to XOR gates, i.e. addition mod 2. We also represented the initial Bell states again with their circuit representation.
\end{enumerate}

\subsection{GHZ state preparation} \label{app:ghzderiv}

\begin{figure}[htb]
	\centering
	\includegraphics[width=0.95\columnwidth]{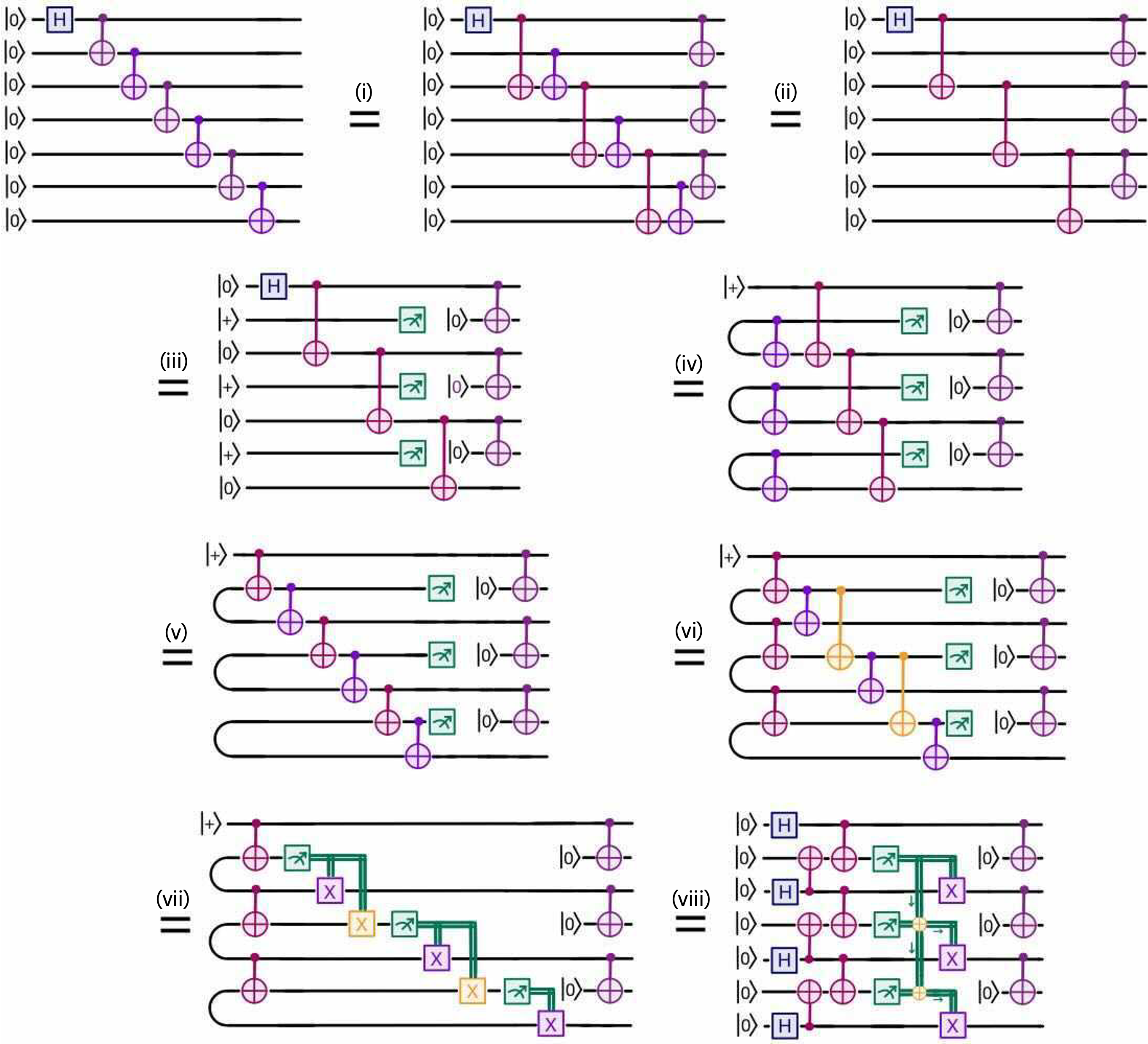}
	\caption{Graphical derivation for the preparation of a GHZ state by converting its canonical but deep unitary circuit into a constant-depth circuit utilizing measurement and feed-forward operations---a dynamic circuit. Roman numerals indicate sequential step numbers described in main text. 
 }
	\label{fig:ghz_derivation}
\end{figure}

In Fig.~\ref{fig:ghz_derivation} we have illustrated a derivation of the GHZ state preparation, exemplary for $7$ qubits, but it can be straightforwardly extended to an arbitrary number of qubits. In the following, we provide explanations for each step of the derivation, labeled by roman numerals in the figure:
\begin{enumerate}[(i)]
\item Pushing every second CNOT gate to the very right introduces the extra pink CNOT gates.
\item We can omit CNOT gates that are conditioned on state $\ket{0}$.
\item As every second qubit is only involved at the very end, we can use those before and reset them.
\item A Bell state followed by a CNOT gate results in two uncorrelated qubits in states $\ket{+}$ and $\ket{0}$.
\item We move the pink CNOT gates along the Bell states to the respective qubits above (they commute with the other CNOTs they are ``passing").
\item Pushing the pink CNOT gates to the left through the purple CNOT gates introduces the extra orange CNOT gates.
\item We make use of the principle of deferred measurement.
\item In a final step we merge the classically-conditioned gates. As the classical calculation can be done extremely fast compared to quantum gates, we draw it as a vertical line. The orange $\oplus$ correspond to XOR gates, i.e. addition mod 2. We also represented the initial Bell states again with their circuit representation.
\end{enumerate}

\section{\mbox{CNOT} circuits} \label{app:comparisoncnot}
\subsection{Unitary variants}
\begin{figure}[htb]
	\centering
	\includegraphics[width=0.95\columnwidth]{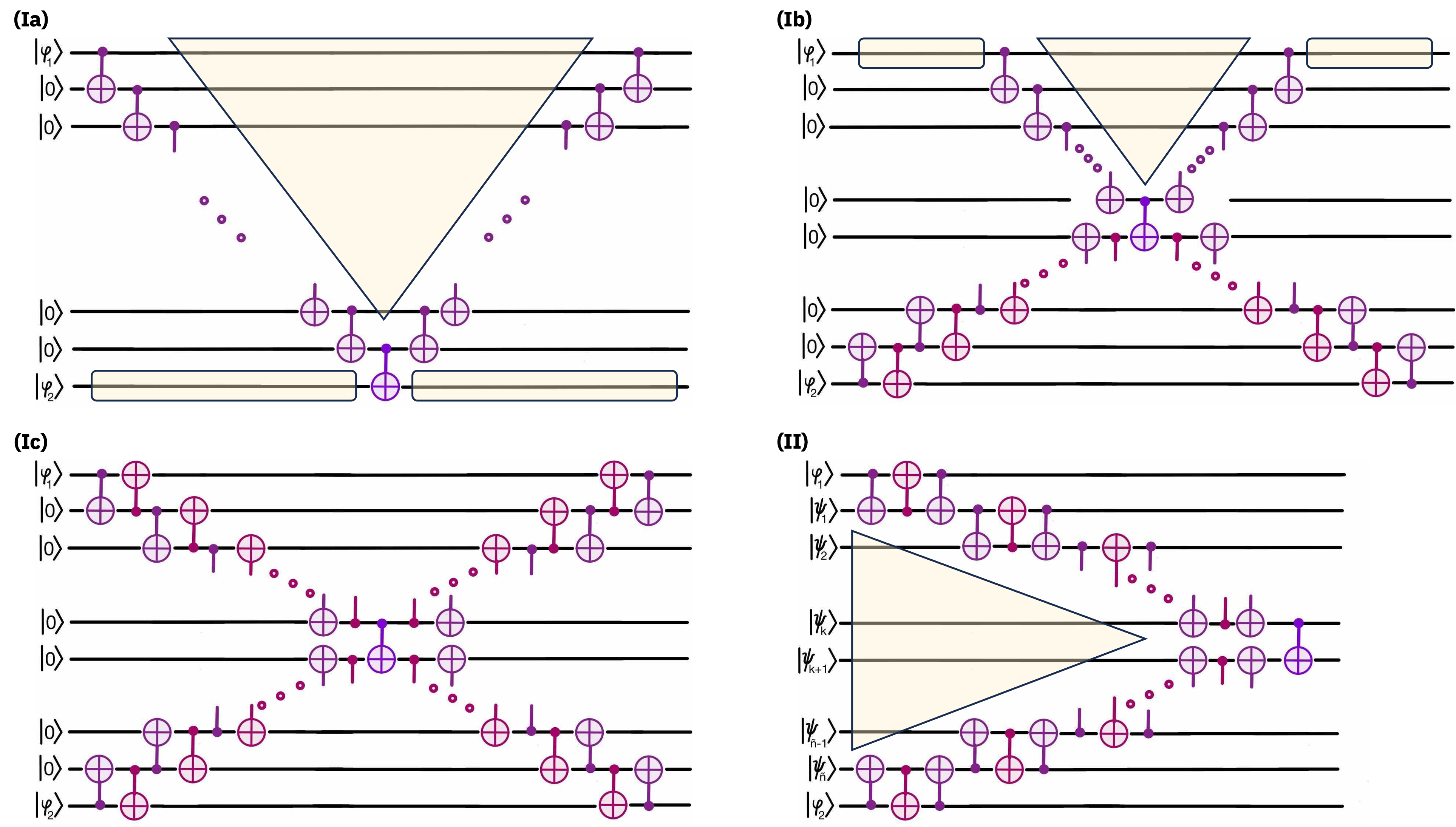}
	\caption{Comparison of the different unitary implementations of a long-range CNOT gate. While the circuits in panels (Ia), (Ib), and (Ic) realize ancilla-based implementations, the circuit of panel (II) realizes a SWAP-based implementation without ancillas. The shaded regions indicate idle periods that accumulate errors.
 }
	\label{fig:comparison_cnots}
\end{figure}

In order to compare the dynamic circuits implementation to a solely unitary one, let us first consider different unitary strategies that might be more or less powerful in different regimes:\\
\\
\textbf{Strategy I: Ancilla-based implementation}\\
We can consider a similar setting as for dynamic circuits, where we place the system qubits in a way that they are connected by a bus of empty ancilla qubits. In this case, we need to swap the system qubits towards each other and back, so that the ancillas are empty in the end again. The swaps can be simplified since the ancillas are empty in the beginning. Here we can divide into different scenarios:
\begin{itemize}
    \item Circuit Ia: To minimize the number of CNOT gates, we could swap the controlled qubit all the way to the target qubit and back, which results in the circuit depicted in Fig.~\ref{fig:comparison_cnots}. Here, a lot of gates cancel, so given $n$ ancilla qubits, the number of CNOT gates is $2n+1$. However, here the idle time of the qubits while they are not in state $\ket{0}$ equals $n^2+2n$ times the CNOT gate time.
    \item Circuit Ib: In order to decrease the idle time, we could essentially swap both, the controlled qubit and the target qubit half-way and back as illustrated in Fig.~\ref{fig:comparison_cnots} (similar to some circuits presented in \cite{kutin2007computation} and \cite{2023arXiv230518128C}). In that case, less gates ``cancel", so for $n$ ancilla qubits we get $3n+1$ CNOT gates, but the idle time reduces to $\frac{n^2}{4}+n$ times the CNOT gate time.
    \item Circuit Ic: If we wanted to reduce the idle time even further, it might be beneficial to not cancel the CNOT gates in scenario 1b), but keep them to bring the swapped qubits back to state $\ket{0}$ as shown in Fig.~\ref{fig:comparison_cnots}. In that case, we have essentially no idle time (as qubits in state $\ket{0}$ are not prone to idling errors). Here, the number of CNOT gates increased to $4n+1$ though.
\end{itemize}
\textbf{Strategy II: SWAP-based implementation without ancillas}\\
This is the case that happens if we just feed our circuit to the transpiler. Here we do not use any ancilla qubits, but only system qubits and apply swaps to move them around. The qubits can be at a different location in the end, so we do not need to swap back. The corresponding circuit is shown in Fig.~\ref{fig:comparison_cnots}. In this case we require $3\tilde{n}+1$ CNOT gates and the idle time is $\frac{3}{2} \tilde{n}^2-2\tilde{n}$ times the CNOT gate time. However, it is important to note here that the number of qubits lying between the two qubits of interest $\tilde{n}$ is on average much shorter than the number of ancillas between two system qubits in the first scenario. Considering the connectivity illustrated in Fig.~\ref{fig:cnot}~(c), the relation is approximately $n = 2\tilde{n}+3$.

\subsection{Error budget} \label{app:cnot_errorbudget}
Let us now compare the regimes in which we expect the different implementations to be most useful to demonstrate the benefit of dynamic circuits. In Appendix~\ref{app:noisemodel} we derive a simple noise model that allows us to compute the combined effect of different sources of decoherence as a single Pauli noise rate:
\begin{align}
\lambda_\mathrm{tot} = t_\mathrm{idle}\lambda_\mathrm{idle} + N_\mathrm{CNOT} \lambda_\mathrm{CNOT} + N_\mathrm{meas} \lambda_\mathrm{meas}\;.\label{eqn:error_tally}
\end{align}
In Lemma~\ref{lemma:product_fidelity} we show that the final process fidelity is loosely lower-bounded by $e^{-\lambda_\mathrm{tot}}$. The quantity $\lambda_\mathrm{tot}$ combines the following noise sources:
\begin{itemize}
\item The total amount of time $t_\mathrm{idle}$ that qubits spend idle within the circuit, and a conversion factor $\lambda_\mathrm{idle}$ that quantifies the strength of decoherence. $t_\mathrm{idle}$ is expressed in multiples of the CNOT gate time (i.e. $t_\mathrm{idle} = 3$ for 3 CNOT gate times). The time for a mid-circuit measurement, including the additional time waiting for feedback, is defined as $\mu$ times the time for a CNOT gate.
\item The total number of CNOT gates $N_\mathrm{CNOTs}$ and an average Pauli noise rate $\lambda_\mathrm{CNOT}$ per CNOT.
\item The total number of mid-circuit measurements $N_\mathrm{meas}$ and an average Pauli noise rate $\lambda_\mathrm{meas}$ per measurement.
\end{itemize}

In Table~\ref{table:cnot_errors}, we have summarized the error budget for each of the cases. 
\begin{table*}[htb]
\begin{ruledtabular}
\begin{tabular}{lcccc}
Case & \begin{tabular}[c]{@{}c@{}}$t_\mathrm{idle}$~\end{tabular} & \begin{tabular}[c]{@{}c@{}}$N_\mathrm{CNOT}$~\end{tabular} & \begin{tabular}[c]{@{}c@{}} $N_\mathrm{meas}$\end{tabular} & \begin{tabular}[c]{@{}c@{}}~Two-qubit gate depth\end{tabular}\\ \hline
Unitary Ia) & $n^2+2n$ & $2n+1$ & $0$ & $2n+1$ \\
Unitary Ib) & $\frac{n^2}{4}+n$ & $3n+1$ & $0$ & $2n+1$ \\
Unitary Ic) & $0$ & $4n+1$ & $0$ & $2n+1$ \\
Unitary II) & $\frac{3}{4}\tilde{n}^2-\frac{3}{2}\tilde{n}$ & $3\tilde{n}+1$ & $0$ & $\frac{3}{2}\tilde{n}+1$ \\
Unitary II) with normed $n$& $\approx\frac{3}{16}n^2-\frac{15}{8}n+\frac{45}{16}$ & $\approx \frac{3}{2}n-2$ & $0$ & $\approx \frac{3}{4}n-\frac{5}{4}$ \\
Dynamic circuits & $2\mu + 2$ & $n+1$ & $n$ & $2+\mu$, or $O(1)$ \\
\end{tabular}
\end{ruledtabular}
\caption{Comparison of the error budget of the unitary and the dynamic circuits implementation in terms of idle time, number of CNOT gates and mid-circuit measurements and two-qubit gate depth. Note, that as the number of involved qubits $\tilde{n}$ needed for the unitary implementation II) is in general much smaller, we rescale it for the error budget with the relation $n\approx 2\tilde{n}+3$}
\label{table:cnot_errors}
\end{table*}

Comparing the different unitary cases it becomes clear that for large $n$ the unitary implementation Ic) will be the best, as all other implementations have an error in the idling time that scales quadratically. This might be slightly counter-intuitive, as it tells us that the extra $2n$ CNOT gates required for implementation Ic) compared to Ia) are worthwhile not being cancelled, as the full swap leaves the other qubits unentangled resulting in a drastically decreased idling error, and as even without measurement and feed-forward, it can be still beneficial to use ancilla qubits and thereby increase the distances. For small $n$ we need to keep in mind that for the swap-based implementation (unitary II) the number of involved qubits $\tilde{n}$ is smaller than the number of qubits $n$ needed for the same task in the ancilla-based implementation. Given the qubit division illustrated in Fig.~\ref{fig:cnot}(d), we achieve a ratio of 31 qubits connected to the bus, 30 qubits not connected to the bus and 66 bus qubits, which is a ratio of roughly 1:2 for all-to-all connected qubits to bus qubits. Respecting the rescaled errors, unitary II would be the most promising implementation for small $n$.
In addition to the CNOT errors and idling errors, for dynamic circuits we also need to consider the error from the additional measurements, as well as a constant term $\mu$ that comes from the idling error during measurement and feed-forward.

\begin{figure}[htb]
	\centering
	\includegraphics[width=0.6\columnwidth]{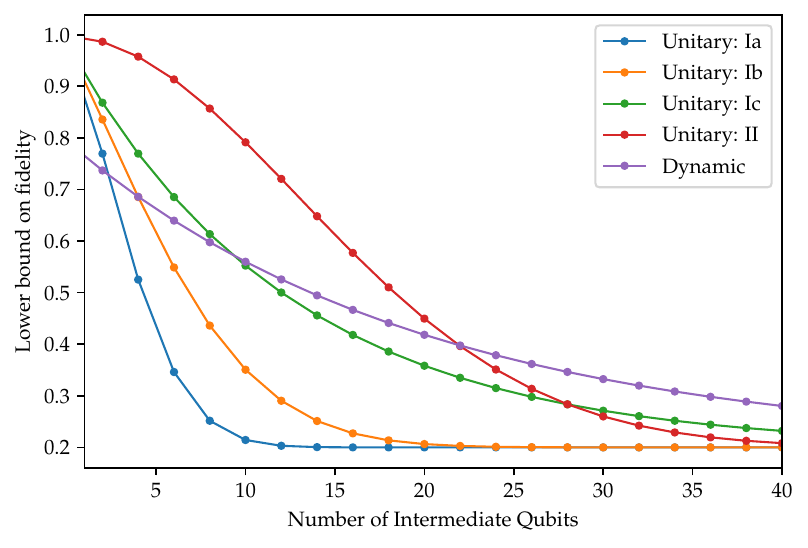}
	\caption{Comparison of the process fidelities of the different unitary implementations as well as the dynamic circuits implementation considering the error budget indicated in Table~\ref{table:cnot_errors}.
 In this figure we use $\mu = (t_{\text{meas}} + t_{\text{feed-forward}})/t_{\text{cnot}} \approx 3.65$, $\lambda_\mathrm{idle}=0.03$, $\lambda_\mathrm{CNOT} =0.02$ and $\lambda_\mathrm{meas}=0.03$.
 }
	\label{fig:fidelities_comparison_cnots}
\end{figure}

Given this rough error analysis in Table~\ref{table:cnot_errors}, we can infer that for large $n$ dynamic circuits will be beneficial if the measurement of $n$ qubits introduces less error than $3n$ CNOT gates, that is, when $\lambda_\mathrm{meas} < 3\lambda_{CNOT}$. A sketch of how the fidelities for the different cases decrease with $n$ is illutrated in Fig.~\ref{fig:fidelities_comparison_cnots}.
Note, that these error analyses only take into account the error on the involved qubits though. Considering also the fact that there are potentially a lot of other $m$ qubits ``waiting" for this operation to be performed would add another idling error of $m\cdot (2n+1)$. So the fact that dynamic circuits can perform entangled gates between arbitrary qubits in constant depth instead of linear depth with only unitary operations speeds up the whole algorithm and therefore might be much more powerful than what we can see in the error on the respective qubits. 

\section{Estimation of the state and gate fidelities using Monte-Carlo sampling}

\subsection{State fidelity} \label{app:statefidelity}
In order to determine the fidelity of the experimentally prepared quantum state, denoted as $\sigma$, we employ the \textit{Monte Carlo state certification} method, which was introduced in Refs.~\cite{daSilva2011,Flammia2011}. We first briefly review the notion of fidelity between two quantum states.

\paragraph*{Quantum state fidelity.}
Let us introduce the Uhlmann-Jozsa state fidelity between two general
quantum states $\rho$ and $\sigma$. These objects are elements of
the space of valid density operators associated with the system Hilbert
space, $\mathcal{H}$, i.e., $\rho,\sigma\in D\left(\mathcal{H}\right)$.
Assuming one of them is a pure state~$\sigma=\kb{\phi}{\phi}$, we
can simplify the general expression as shown in the following: 
\begin{align}
\label{eq:UhlmanFidelity}
F\left(\rho,\sigma\right) & \isdef\left[\Tr\left(\sqrt{\sqrt{\rho}\sigma\sqrt{\rho}}\right)\right]^{2}\\
 & =\braOket{\phi}{\rho}{\phi}\\
 & =\Tr\left[\rho\sigma\right]\;.
\end{align}
If $\rho$ is also a pure state $\rho=\kb{\psi}{\psi}$, the expression
reduces to a simple overlap $F\left(\rho,\sigma\right)=\left|\braket{\psi}{\phi}\right|^{2}$. We note that some authors define the square root of this as the fidelity.

\paragraph*{Pauli decomposition.}
To connect to experimental measurements, let us decompose the quantum
sates in the standard Pauli basis. The set of all Pauli operators
on $n$ qubits $\left\{ I,X,Y,Z\right\} ^{\otimes n}$ forms an orthogonal
Hermitian operator basis. The inner product in operator space $L\left(\mathcal{H}\right)$
between two Pauli operators $P_{i},P_{j}\in\mathcal{L}\left(\mathcal{H}\right)$
is $\left\langle P_{i},P_{j}\right\rangle =\Tr\left(P_{i}P_{j}\right)=d\delta_{ij}$,
where the dimension of the pure state Hilbert space $d\isdef\dim\mathcal{H}=2^{n}$.
In terms of this basis, any quantum state $\rho\in D\left(\mathcal{H}\right)$,
can be decomposed into 
\[
\rho=\sum_{i=0}^{4^{n}-1}\frac{\left\langle P_{i},\rho\right\rangle }{\left\langle P_{i},P_{i}\right\rangle }P_{i}=\frac{1}{d}\sum_{i=0}^{4^{n}-1}\rho_{i}P_{i}\;,\quad\mathrm{with}\quad\rho_{i}\isdef\left\langle P_{i},\rho\right\rangle =\Tr\left(P_{i}\rho\right)\;,
\]
where the Pauli expectation value of the state with respect to the $i$-th Pauli is $\rho_{i}$---an easily measurable quantity.
We can similarly define the expectation values of the Pauli~$P_i$ with respect to the prepared state $\sigma$ and the desired state $\rho$ as $\sigma_i := \langle P_i \rangle_\sigma = \tr \left(\sigma P_i\right)$ and $\rho_i := \langle P_i \rangle_\rho = \tr \left(\rho P_i \right)$, respectively.

\paragraph*{Fidelity in terms of Pauli expectation values.}
The state fidelity between the measured~$\sigma$ and ideally expected pure~$\rho$ state, see Eq.~\eqref{eq:UhlmanFidelity},  in terms of the Pauli decomposition of each is 
\begin{align}
    F(\rho,\sigma) = \tr \left[\rho \sigma\right] = \sum_i \frac{\rho_i \sigma_i}{d}
    =\sum_i \frac{\rho_i^2}{d} \frac{\sigma_i}{\rho_i}\;,
\end{align}
where $\sigma_i$ is an experimentally measured expectation value and  $\rho_i$ is a  theoretically calculated one. Given this, we can now define the \textit{relevance distribution} $r(P_i):=\frac{\rho_i^2}{d}$, such that $F(\rho,\sigma) = \sum_{i: \rho_i \neq 0} r(P_i) \frac{\sigma_i}{\rho_i}$.

\paragraph*{Random sampling of expectation values.}
When sampling $m$ random operators $\{P_{k}\}_{k=1..m}$ according to the relevance distribution $r(P_k)$ and determining their expectation values $\sigma_{k}$, the estimated fidelity $\tilde{F} := \sum_{k=1}^m \frac{\sigma_{k}}{\rho_{k}}$ approximates the actual fidelity $F$ with an uncertainty that decreases as $\frac{1}{\sqrt{m}}$.
Note that there is also an uncertainty in estimating each $\sigma_{k}$, where for an additive precision~$\epsilon$ roughly $(\epsilon \rho_{k})^{-2}$ shots are required.

\paragraph*{Random sampling of GHZ  stabilizers.}
As the GHZ state is a stabilizer state, for each $n$ there are exactly $2^n$ non-zero Pauli operators $P_i$ that each have eigenvalue $\pm 1$. Note that some stabilizers of the GHZ state have a minus sign, \textit{e.g}. $-YYX$. For the $n$-qubit GHZ state, by defining the set of stabilizers $\{S_i\}_{i = 1..2^n}$, we can express the fidelity in terms of only expectation values on the stabilizers
\begin{align}
    F(\rho,\sigma) &= \frac{1}{2^n}\sum_{i=1}^{2^n} \langle S_i \rangle_\sigma\;.
\end{align}
This expression can be approximated by randomly sampling~$m$ of the $2^n$ stabilizers, defining the unbiased estimator~$\tilde{F} = \frac{1}{m}\sum_{k=1}^m \langle S_k \rangle_\sigma = F + \mathcal{O}\left(\frac{1}{\sqrt{m}}\right)$, which converges with the number of random samples chosen to the ideal fidelity.

\subsection{Average gate fidelity} \label{app:gatefidelity}

Similarly to the state fidelity, we use the \textit{Monte Carlo process certification} following \cite{daSilva2011} to determine the average gate fidelity of our noisy CNOT gate.

\paragraph*{Average gate fidelity.} 
Consider the case in which we want to implement an ideal gate~$\mathcal{U}(\rho)\isdef U\rho U^{\dagger}$.
However, instead we can implement only a noisy gate~$\tilde{\mathcal{U}}(\rho)\isdef\mathcal{U}(\Lambda(\rho))$, where $\Lambda$ is some effective noise channel and $\rho$ is a quantum state. What is the gate
fidelity of noisy~$\tilde{\mathcal{U}}$ relative to the ideal~$\mathcal{U}$? For a single given pure state $\rho=\kb{\phi}{\phi}$, the state fidelity of the output of the ideal and noisy channels is
\begin{align}
F\left(\mathcal{U},\tilde{\mathcal{U}};\rho\right)&=\left[\Tr\left[\sqrt{\sqrt{\mathcal{U}\left(\rho\right)}\tilde{\mathcal{U}}\left(\rho\right)\sqrt{\mathcal{U}\left(\rho\right)}}\right]\right]^{2}\\
&=\Tr\left[\mathcal{U}\left(\rho\right)\tilde{\mathcal{U}}\left(\rho\right)\right]\\
&=\Tr\left[\rho\Lambda\left(\rho\right)\right]\;,
\end{align}
which can be used to obtain the average gate fidelity devised by a uniform Haar average over the fidelity of the ideal and noisy output states, with $\rho_{\psi}=\kb{\psi}{\psi}$,
\begin{align}
    \mathcal{F}_{\mathrm{avg}}\left(\mathcal{U},\tilde{\mathcal{U}}\right) 
    &= \int\mathrm{d}\psi\,F\left(\mathcal{U},\tilde{\mathcal{U}};\rho_{\psi}\right)
    \\
    &= \int\mathrm{d}\psi\,\Tr\left[\mathcal{U}\left(\rho\right)\tilde{\mathcal{U}}\left(\rho\right)\right]\\
    &= \Tr\left[\int\mathrm{d}\psi\,\kb{\psi}{\psi}\Lambda\left(\kb{\psi}{\psi}\right)\right]\;.
\end{align}
To estimate~$\mathcal{F}_{\mathrm{avg}}\left(\mathcal{U},\tilde{\mathcal{U}}\right)$, we will use the process (or entanglement) fidelity as a more experimentally-accessible quantity.

\paragraph*{Process fidelity.} 
Compared to the gate fidelity, the process fidelity is more readily estimated. It can in turn serve as a direct proxy to the gate fidelity. To make the connection, recall that the Choi-Jamiolkowski isomorphism \cite{Jamiolkowski1972} maps every quantum operation $\Lambda$ on a $d$-dimensional space to a density operator $\rho_\Lambda = (\mathbb{I}\otimes\Lambda)\ket{\phi}\bra{\phi}$, where $\ket{\phi} = \frac{1}{\sqrt{d}}\sum_{i=1}^d\ket{i}\otimes\ket{i}$. 
For a noise-free, ideal unitary channel~$\mathcal{U}$ and its experimental, noisy implementation~$\tilde{\mathcal{U}}$, the process fidelity $\mathcal{F}_\mathrm{proc}$ is the state fidelity of the respective Choi states $\rho_{\mathcal{U}}$ and $\rho_{\tilde{\mathcal{U}}}$:
\begin{align}
\mathcal{F}_\mathrm{proc}
({{\mathcal{U}}}, {\tilde{\mathcal{U}}})
:= 
F(\rho_{{\mathcal{U}}}, \rho_{\tilde{\mathcal{U}}}) \;.
\end{align}
From this fidelity, the gate fidelity can be extracted using the following relation derived in \cite{Horodecki1999}:
\begin{align}
\mathcal{F}_\mathrm{gate}({{\mathcal{U}}}, {\tilde{\mathcal{U}}})
= 
\frac{d \mathcal{F}_\mathrm{proc}(\rho_{{\mathcal{U}}}, \rho_{\tilde{\mathcal{U}}})+1}{d+1}\;.
\end{align}

\paragraph*{Estimating the process fidelity.} As described in Ref.~\cite{daSilva2011}, instead of a direct implementation of $(\mathbb{I}\otimes\tilde{\mathcal{U}})\ket{\phi}\bra{\phi}$ followed by measuring random Pauli operators on all qubits, we follow the more practical approach, where $\tilde{\mathcal{U}}$ is applied to the complex conjugate of a random product of eigenstates of local Pauli operators $P_i \otimes P_j$, followed by a measurement of random Pauli operators $P_k \otimes P_l$. This leads to the same expectation values 
\begin{align}\rho_{ijkl}:=\tr\left[(P_i \otimes P_j \otimes P_k \otimes P_l)(\mathbb{I}\otimes \mathcal{U})\ket{\phi}\bra{\phi}\right]= \tr\left[(P_k \otimes P_l) \mathcal{U} (P_i \otimes P_j)^\ast\right]/d\;.\end{align} 
The operators are then sampled according to the relevance distribution 
\begin{align}r_{ijkl}:=r(P_iP_jP_kP_l)=\frac{\rho_{ijkl}^2}{\tilde{d}}\;.\end{align}
Note, that $\tilde{d}$ corresponds to the dimension of the Choi state, i.e. here $\tilde{d}=16$. For $\Lambda(\rho) =\mathrm{CNOT} \rho \mathrm{CNOT}^\dagger$, there are only $16$ combinations of Pauli operators with a non-zero expectation value $\rho_{ijkl}$:  $\rho_{ijkl}=-1$ for $P_iP_jP_kP_l \in \ \{YYXZ,XZYY\}$ and $\rho_{ijkl}=1$ for the remaining 14. Thus, the relevance distribution is uniform amongst those with $r=\frac{1}{16}$ and we can just take the average expectation value of those $16$ operators.

\section{Experimental details}
\label{app:expdetails}
\subsection{CNOT gate teleportation}
\label{app:expdetails_cnot}
\begin{figure*}[htb]
\includegraphics[width=1.0\columnwidth]{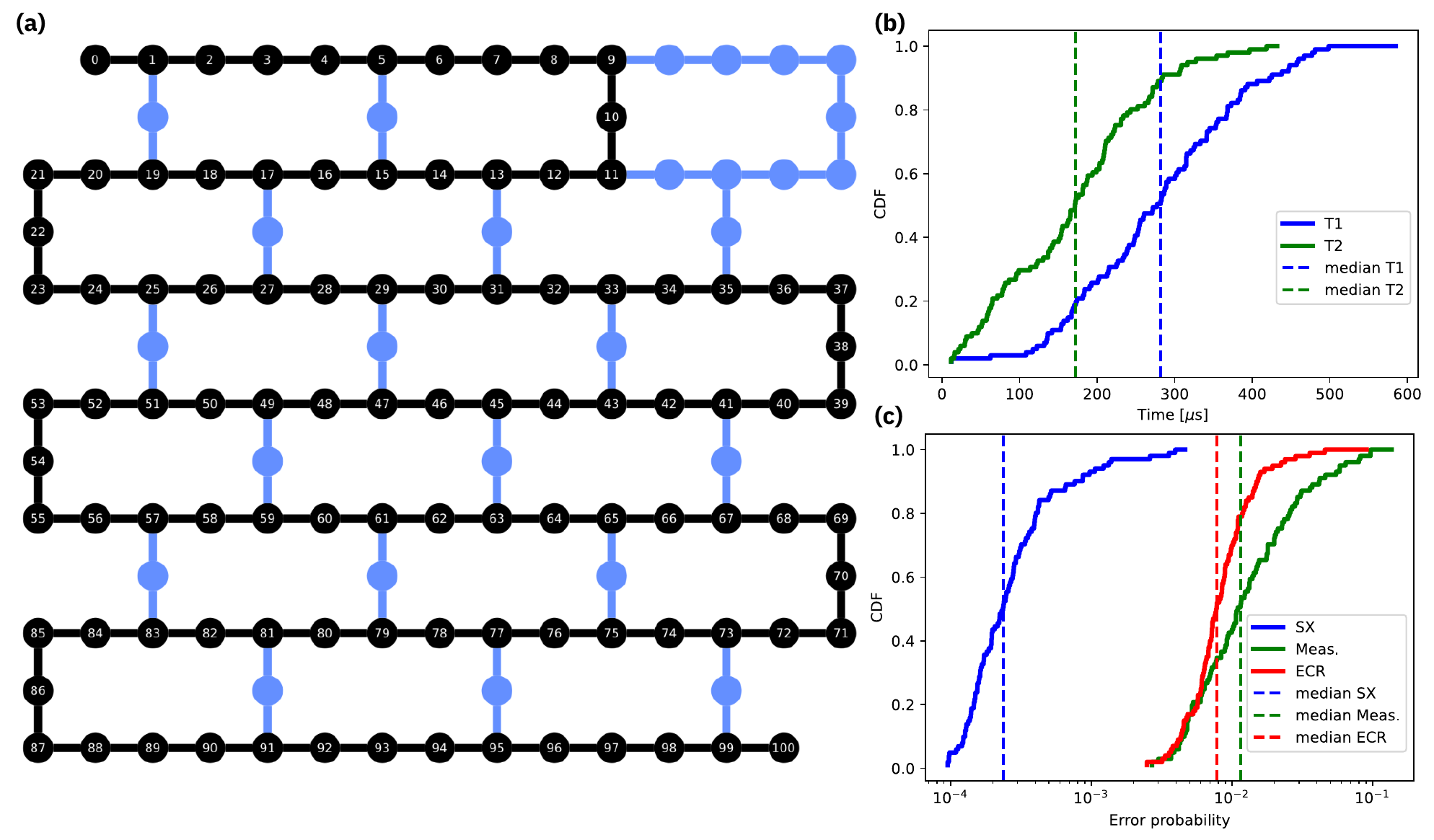}
\caption{Implementation details. In (a), we show the device layout of \texttt{ibm\_sherbrooke}, with the 101 qubits chosen for our dynamic circuits marked in black. In (b) and (c), we plot the cumulative distribution of the T1 and T2 coherence times, the single qubit gate (SX), readout (Meas.) and two qubit echoed cross-resonance gate (ECR) error rates of the chosen qubits, as well as the corresponding median values.}
\label{fig:qubitdata}
\end{figure*}

We perform the long-range gate teleportation experiments on \texttt{ibm\_sherbrooke}, a 127-qubit superconducting quantum processor. The line of 101 qubits chosen for the experiments are indicated in \cref{fig:qubitdata}(a). The cumulative distribution of their T1 and T2 coherence times, as well as of their different error rates are shown in \cref{fig:qubitdata}(b)-(c), indicating also the corresponding median values. 
The two-qubit gate time is $0.5~\mu s$, the readout time $1.2~\mu s$ and the feed-forward time roughly $0.7~\mu s$.

\subsection{GHZ state preparation}
\label{app:expdetails_ghz}
\begin{figure*}[htb]
\includegraphics[width=0.8\columnwidth]{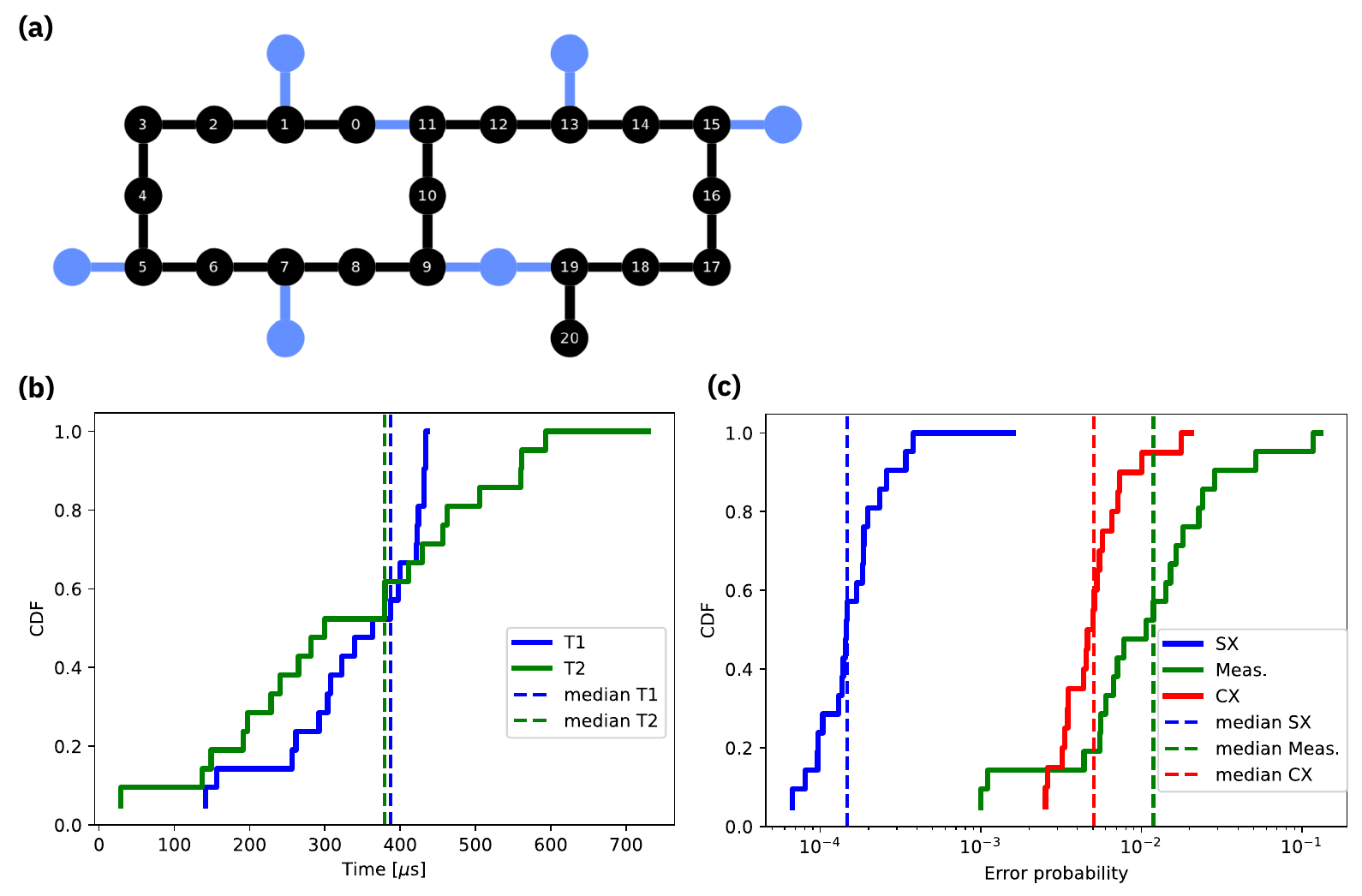}
\caption{Implementation details. In (a), we show the device layout of \texttt{ibm\_peekskill}, with the 21 qubits chosen for our dynamic circuits marked in black. In (b) and (c), we plot the cumulative distribution of the T1 and T2 coherence times, the single qubit gate (SX), readout (Meas.) and two qubit controlled-X (CX) error rates of the chosen qubits, as well as the corresponding median values.}
\label{fig:qubitdata_ghz}
\end{figure*}

We perform the GHZ state preparation experiments on \texttt{ibm\_peekskill}, a 27-qubit superconducting quantum processor. The line of 21 qubits chosen for the experiments are indicated in \cref{fig:qubitdata_ghz}(a). The cumulative distribution of their T1 and T2 coherence times, as well as of their different error rates are shown in \cref{fig:qubitdata_ghz}(b)-(c), indicating also the corresponding median values.
The two-qubit gate time is $0.6~\mu s$, the readout time $0.9~\mu s$ and the feed-forward time roughly $0.7~\mu s$.

\section{\mbox{Toffoli} or \mbox{CCZ}}\label{app:toffoli}
\begin{figure}[htb]
    \centering    \includegraphics[width=0.7\columnwidth]{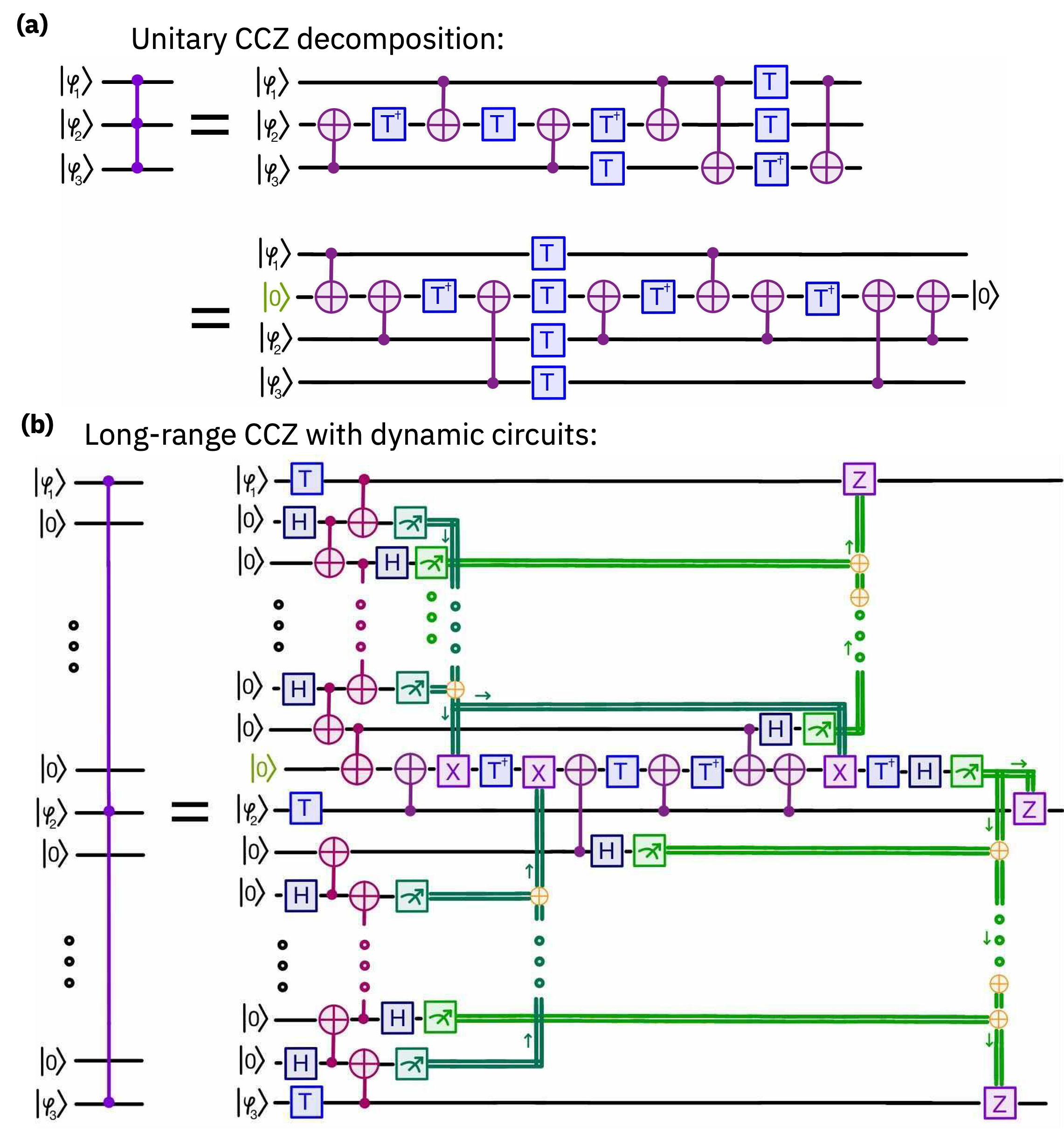}
    \caption{\mbox{CCZ} with (a) unitary circuit and (b) a dynamic circuit over long ranges.
    }
    \label{fig:ccz_circuit}
\end{figure}

Dynamic circuits can also be applied to more efficiently compile multi-qubit gates. As an example, we describe how the \mbox{CCZ}, or \mbox{Toffoli} gate up to two single-qubit Hadamard gates, can be implemented by optimizing multiple teleported \mbox{CNOT} gates. Compilation of the unitary circuit on a 1D chain of $n+3$ qubits using \mbox{CNOT} gates na\"ively requires a two-qubit gate depth of $\mathcal{O}\left(n\right)$.
Using dynamic circuits, we can implement this long-range entangling gate in shallow depth.
Na\"ively, one could successively implement each \mbox{CNOT} gate of the typical Toffoli decomposition (shown at the top of Fig.~\ref{fig:ccz_circuit}(a)) using the gate teleportation described previously. However, involving an ancillary qubit between the three system qubits to merge the teleported gates, as shown at the bottom of Fig.~\ref{fig:ccz_circuit}(a), allows for a more efficient implementation with the dynamic circuit; see Fig.~\ref{fig:ccz_circuit}(b).
In total, this formulation requires $n+1$ measurements, $n+6$ \mbox{CNOT} gates, and 5 feed-forward operations divided across two sequential steps.
Notably, as most qubits are projectively measured early in the circuit, the idling error should be low. Thus, we expect this shallow implementation with dynamic circuits to be advantageous over its unitary counterpart, especially for large $n$.

\section{Error analysis for GHZ states} 
\subsection{Error budget}
\label{app:GHZerrorbudget}

As in Appendix~\ref{app:cnot_errorbudget}, we leverage (\ref{eqn:error_tally}) to estimate the total noise $\lambda_\mathrm{tot}$ of a quantum circuit as motivated by the model discussed in Appendix~\ref{app:noisemodel}. There, it is derived that $e^{-\lambda_\mathrm{tot}}$ gives a lower bound on the \emph{process} fidelity of the circuit. For GHZ states however, we are interested the \emph{state} fidelity, so the bound from Lemma~\ref{lemma:product_fidelity} no longer applies in a rigorous sense. However, we find that the same model can still provide useful intuition if we accept that the model parameters $\lambda_\mathrm{CNOT}, \lambda_\mathrm{meas}$ no longer have a direct interpretation in terms of worst-case Pauli-Lindblad noise or a combination of amplitude- and phase-damping noise respectively. See Appendix~\ref{app:noisemodel} for details.

For the unitary approach we require $n$ CNOT gates to entangle $n$ qubits. For simplicity we assume (and implement) only a one-dimensional connectivity chain in our protocols and the following numbers correspond to an even number $n$ (only constant terms change when considering odd $n$). To minimize the idling time, we start in the middle and apply CNOT gates simultaneously towards both ends. This leads to an idle time of $\frac{n^2}{4}-\frac{3}{2}n+2$ times the CNOT gate time, as displayed in Table~\ref{table:ghz_errors}.
In the dynamic circuits approach we require $\frac{3}{2}n-2$ CNOT gates in total, while the idling time is $\frac{\mu}{2}n+1$ times the CNOT gate time, where $\mu$ corresponds to the measurement and feed-forward time (as a multiple of the CNOT gate time). However, here we also need to consider the errors of the additional $\frac{n}{2}-1$ measurements. As the error coming from the CNOT gates and the measurements is usually substantially larger than the error from the idling time, we expect that for small $n$ the standard unitary preparation succeeds. However, as the idling time there scales as $\mathcal{O}\left(n^2\right)$ in contrast to all errors in the measurement-based approach scaling only as $\mathcal{O}\left(n\right)$, we expect a crossover for large $n$, where the implementation with dynamic circuits will become more beneficial. The error budget is summarized in Table~\ref{table:ghz_errors}. 

\begin{table*}[htb]
\begin{ruledtabular}
\begin{tabular}{lcccc}
Case & \begin{tabular}[c]{@{}c@{}}$t_\mathrm{idle}$~\end{tabular} & \begin{tabular}[c]{@{}c@{}} $N_\mathrm{CNOT}$~\end{tabular} & \begin{tabular}[c]{@{}c@{}} $N_\mathrm{meas}$\end{tabular} & \begin{tabular}[c]{@{}c@{}}~Two-qubit gate depth\end{tabular}\\ \hline
Unitary & $n^2/4-3n/2+2$ & $n-1$ & $0$ & $n-1$ \\
Dynamic circuits & $1+\mu n /2$ & $3n/2-2$ & $n/2-1$ & $3+\mu$, or $O(1)$ \\
\end{tabular}
\end{ruledtabular}
\caption{Comparison of the error budget of the unitary and the dynamic circuits implementation in terms of idle time, number of CNOT gates and mid-circuit measurements and two-qubit gate depth.}
\label{table:ghz_errors}
\end{table*}

\subsection{Expected cross-over for lower mid-circuit measurement errors} \label{app:crossover}
\begin{figure}[htb]
	\centering
	\includegraphics[width=0.8\columnwidth]{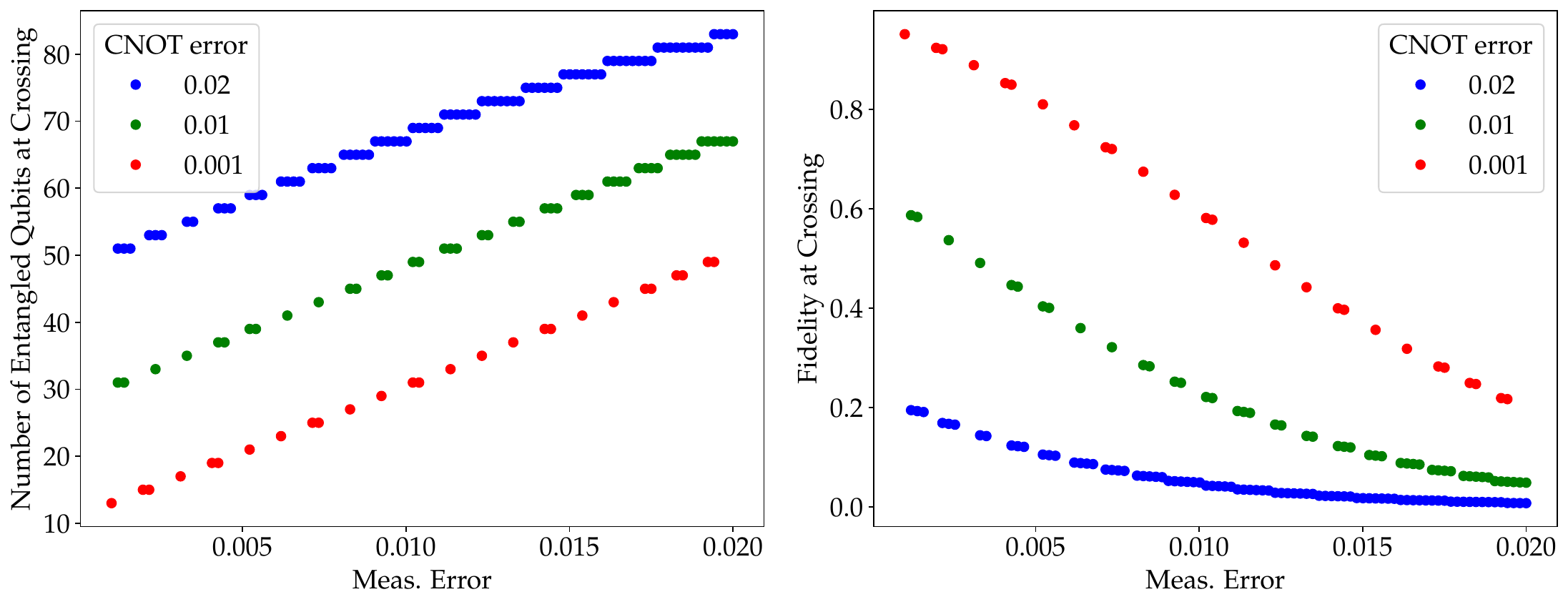}
	\caption{Noise-model predictions that indicate how many qubits are required to see a cross-over and what the corresponding fidelity would be as a function of the mid-circuit measurement errors.
 }
 \label{fig:crossover}
\end{figure}

In Fig.~\ref{fig:crossover} we determine the expected crossover in performance from unitary to dynamic circuits for varying mid-circuit measurement and CNOT gate errors.
We use the values of $t_\mathrm{idle}, N_\mathrm{CNOT},$ and $N_\mathrm{meas}$ shown in Table~\ref{table:ghz_errors} to predict how many qubits are required to see and the state fidelity at the cross-over, or where the performance of dynamic circuits becomes higher than that of its unitary counterpart, as a function of the mid-circuit measurement errors. Note that in this noise model we assume that we can eliminate all ZZ errors by applying dynamical decoupling. We keep the idling error constant at $\lambda_\mathrm{idle} = 0.001$ and consider different CNOT errors $\lambda_\mathrm{CNOT} \in \{0.001, 0.01, 0.02\}$. We can reach a fidelity $>0.5$ for a CNOT error of $\lambda_\mathrm{CNOT} = 0.01$ with mid-circuit measurement errors $\lambda_\mathrm{meas}\lesssim0.003$ and for a CNOT error $\lambda_\mathrm{CNOT} = 0.001$ with mid-circuit measurement errors $\lambda_\mathrm{meas}\lesssim0.012$

\section{Pauli-Lindblad Noise Model} \label{app:noisemodel}

In this section we present a simple framework for computing lower bounds on fidelities using the  Pauli-Lindblad noise model discussed in \cite{Berg2022}. Pauli-Lindblad noise channels have several nice properties that we can use to simplify calculations, and also allow us to reduce estimates of the noise properties of our hardware to relatively few parameters.

Normally, Pauli-Lindblad noise is the workhorse of probabilistic error cancellation - an error mitigation scheme that leverages characterization of noise in order to trade systematic uncertainty for statistical uncertainty. But we are more interested in using Pauli-Lindblad noise as a tool for capturing the behavior of fidelity as a function of circuit size with an appropriate balance of rigor and simplicity.

As such, our central goal in this section is to develop mathematical tools that allow us develop a Pauli-Lindblad representation of various noise sources such as decoherence and gate noise and to find a method to combine all of this noise into a fidelity for the entire process. In particular, we aim to give a justification for modeling noise via the quantity $\lambda_\mathrm{tot}$ as in (\ref{eqn:error_tally}). This is achieved by Lemma~\ref{lemma:product_fidelity}, which states that $e^{-\lambda_\mathrm{tot}}$ gives a lower bound on the process fidelity.

We leave the majority of our mathematical exposition without proof for sake of brevity, but present the proof of Lemma~\ref{lemma:product_fidelity} at the end of this section.

\paragraph*{Pauli-Lindblad noise.}  Pauli-Lindblad noise is a quantum channel defined as follows. Let $\mathcal{P}$ be the $n$-qubit Pauli group modulo phase, and consider some $P \in \mathcal{P}$. Then for some noise rate $\lambda \in \mathbb{R}^+$ the noise channel $\Gamma^\lambda_P$ is given by:
    \begin{align}
        \Gamma_P^\lambda(\rho) = (1-\omega) \rho + \omega P\rho P^\dagger \quad \mathrm{where}\quad \omega := \frac{1-e^{-2\lambda}}{2}\;.
    \end{align}

This is essentially applying $P$ with probability $\omega$. Pauli noise channels also have a representation as time evolution with respect to a simple Lindbladian: for $P \in \mathcal{P}$, let $\mathcal{L}_P(\rho) := P \rho P - \rho$. This way $\Gamma_P^\lambda = e^{\lambda \mathcal{L}_P}$. 

The main justification for why we can restrict to Pauli noise channels is twirling. Conjugating an arbitrary noise channel by a random Pauli matrix yields a channel that is always expressible as a product of Pauli noise.  Although our experiments do not feature twirling, even for untwirled circuits we expect the Pauli-Lindblad noise to capture the first-order noise behavior.

Another reason why we expect our noise model to only capture the behavior to first-order is that we assume the noise rates are the same for all qubits. All CNOT gates and idle times are assumed to contribute the same amount of noise. But this is not a realistic representation of our hardware - in actuality different qubits have different coherence times and gate qualities also vary. When we consider circuits on many qubits we expect these differences to average out.

 Let~$\Lambda$ be a quantum channel. Then let $\tilde \Lambda$ be its Pauli-twirled version given by:
    \begin{align}
    \tilde\Lambda := \frac{1}{|\mathcal{P}|} \sum_{P \in \mathcal{P} } P \Lambda(P \rho P) P\;.
    \end{align}
For $Q \in \mathcal{P}$, twirled channels $\tilde \Lambda$ satisfy $\tilde \Lambda(Q) = c_Q Q$ for some coefficients $c_Q$. For every $\tilde \Lambda$ there exist noise rates $\lambda_P$ for $P \in \mathcal{P}/\{I\}$ such that $\tilde\Lambda = \prod_P \Gamma_P^{\lambda_P}$. These noise rates satisfy:
    \begin{align}
        c_Q = e^{ -2 \sum_{P} ( \lambda_P \cdot 1_{PQ=-QP}) }\;.
    \end{align}

A central convenience of Pauli noise channels is that they do not interfere with each other when propagated: Pauli noise channels commute $\Gamma^{\lambda_P}_P\Gamma^{\lambda_Q}_Q = \Gamma^{\lambda_Q}_Q\Gamma^{\lambda_P}_P$, and the noise rates can be added together when the Pauli is the same $\Gamma^{\lambda_1}_P\Gamma^{\lambda_2}_P = \Gamma^{\lambda_1+\lambda_2}_P $\;.

\paragraph*{Combining noise channels into a single fidelity.} Say we are trying to compute the overall amount of noise in a particular quantum circuit that has been appropriately twirled. Gates and idle time of the qubits all contribute some amount of Pauli noise. We propagate all of the Pauli noise to the end of the circuit, thereby removing any noise that does not affect certain mid-circuit measurements. Finally, we must tally up the noise Paulis on the resulting quantum state.

One metric for measuring the error on the final state is trace distance, or diamond norm if we are considering a channel. For a single Pauli noise source, we have the simple relation that for any $P$ we have $\left|\Gamma^\lambda_P - I\right|_\diamond = 1-e^{-2\lambda}$.  To generalize this to multiple Paulis, a simple approach could be to just apply the triangle inequality to all of the different Paulis. But it turns out we can do much better using the following bound on the process fidelity:

\begin{lemma} \label{lemma:product_fidelity} Consider a channel $\Lambda = \prod_{P} \Gamma_P^{\lambda_P}$ for some rates $\lambda_P$. Then $\mathcal{F}_\mathrm{proc}(\Lambda, \mathcal{I}) \geq \exp(-\sum_P \lambda_P)$.
\end{lemma}

This bound is still pretty loose, but it is very simple and does better than adding up diamond norms. This can be seen by, for example, looking at the channel $\prod_{i=1}^N \Gamma_{P_i}^{c/N}$. Lemma~\ref{lemma:product_fidelity} gives $\mathcal{F}_\mathrm{proc} \geq \exp(-c)$ while adding up diamond norms and converting them to a fidelity bound gives $\mathcal{F}_\mathrm{proc} \geq 1 - \frac{1}{2}N(1-e^{-2c/N})$. The latter is looser for $N \geq 2$ and for any $c$. 

Lemma~\ref{lemma:product_fidelity} also has the key advantage that it makes computation of the overall noise rate very simple: just add up all the noise rates. This allows us to simply tally the total idle time and count the number of CNOTs to obtain the total amount of noise, as in Appendix~\ref{app:cnot_errorbudget}.

An issue with using Lemma~\ref{lemma:product_fidelity} is that it becomes increasingly loose in the limit of large $\sum_P \lambda_P$. The quantity $\exp(-\sum_P \lambda_P)$ vanishes in this limit, but in general we have $\mathcal{F}_\mathrm{proc}(\Lambda, \Lambda') \geq 1/d$ for all $\Lambda,\Lambda'$. When we only have one source of Pauli noise $\Gamma_P^{\lambda}$ then not even the lower limit of $1/d$ can be reached as $\lambda \to \infty$. Unfortunately, we see no way of overcoming this limitation while preserving the mathematical elegance of this tool: we would like to simply consider the quantity $\sum_P \lambda_P$. The reason for this shortcoming is that we do not account for cancellations between Pauli errors - we discuss the details of the derivation at the end of this section.

Another limitation of this analysis is that it completely ignores crosstalk. Every gate is assumed to behave independently. Assuming independent errors corresponds to a worst-case analysis analogous to the union bound, so we would expect the bounds resulting from Lemma~\ref{lemma:product_fidelity} to still roughly capture average error from crosstalk by accounting for it as $T_2$ dephasing noise, an error that we include when modeling experiments without dynamical decoupling.

\paragraph*{Propagating noise to the end of the circuit.} Next, we discuss how to move all the noise sources to the end of the circuit. This is particularly easy since we are considering Clifford circuits. Once all the noise is in one place, we can use Lemma~\ref{lemma:product_fidelity} to combine it into a single fidelity. 

With $\mathcal{U}\cdot\isdef U\cdot U^{\dagger}$ as before, elementary calculation shows that $\mathcal{U}\Gamma^\lambda_P = \Gamma^\lambda_{\mathcal{U}(P)} \mathcal{U}$, so Pauli-Lindblad noise propagated through a unitary Clifford circuit is still Pauli-Lindblad noise.

Our circuits also feature several adaptive gates, propagation through which can be achieved as follows.  Let $\Lambda_\mathrm{disc}$ be the channel that traces out the first of two qubits. Then $\Lambda_\mathrm{disc} \Gamma^\lambda_{P\otimes Q} = \Gamma^\lambda_{Q} \Lambda_\mathrm{disc}$. Similarly, let $\Lambda_{\mathrm{corr},P}$ be the channel that measures the first qubit and applies a correction $P$ onto the second qubit. If $P$ and $Q$ commute, then $\Lambda_{\mathrm{corr},P}\Gamma^\lambda_{Q\otimes R}=\Gamma^\lambda_{R}\Lambda_{\mathrm{corr},P}$. Otherwise, $\Lambda_{\mathrm{corr},P}\Gamma^\lambda_{Q\otimes R}=\Gamma^\lambda_{PR}\Lambda_{\mathrm{corr},P}$.

Now that we have established how to move noise to the end of the circuit and to tally it into a bound on the fidelity, all that remains is to show how to bring various noise sources into Pauli-Lindblad form.

\paragraph*{Decoherence noise.} We begin with decoherence noise that affects idling qubits. We consider depolarizing, dephasing, and amplitude damping noise.

Conveniently, depolarizing and dephasing noise are already Pauli noise channels. A depolarizing channel $\Lambda_{\mathrm{dep},q}$ replaces the input $\rho$ with the maximally mixed state with probability $1-q$:
    \begin{align}
        \Lambda_{\mathrm{dep},q}(\rho)  = q\rho + (1-q) \frac{I}{2^n}\;.
    \end{align} 
 We derive that $\Lambda_{\mathrm{dep},q} =  \prod_{P \in \mathcal{P}_n/\{I\}} \Gamma^\lambda_P$ with $q = \exp(-4^n\lambda)$.

 The phase damping process is given by the Lindbladian with $L_0 = \ket{0}\bra{0}$ and $L_1 = \ket{1}\bra{1}$:
    \begin{align}
        \mathcal{L}_\mathrm{ph} = \sum_{i\in\{0,1\}} L_i \rho L_i^\dagger - \frac{1}{2}\left\{L^\dagger_i L_i , \rho\right\}\;.
    \end{align}
Since $\mathcal{L}_\mathrm{ph} = \frac{1}{2} \left(Z\rho Z - \rho\right)$, it satisfies $e^{\lambda \mathcal{L}_\mathrm{ph}} = \Gamma^{\lambda/2}_Z$. We can easily compute $\lambda$ from a phase damping experiment: since $\bra{+} \Lambda^\lambda_\mathrm{damp}(\ket{+}\bra{+})\ket{+} = \frac{1}{2}( 1 + e^{-\lambda})$ we have $\lambda = t/T_2$.

The amplitude damping channel is not a Pauli-Lindblad channel, and must be twirled in order to bring into Pauli-Lindblad form.  The amplitude damping process $\mathcal{L}_\mathrm{damp}$ is given by $L = \ket{0}\bra{1}$ with:
    \begin{align}
\mathcal{L}_\mathrm{damp}(\rho) = L\rho L - \frac{1}{2}\left\{L^\dagger L ,\rho\right\}\;.
    \end{align}
    If we let $\Lambda^\lambda_\mathrm{damp} := e^{\lambda \mathcal{L}_\mathrm{damp}}$ then we have $\tilde \Lambda^\lambda_\mathrm{damp} = \Gamma^{\lambda/4}_X \Gamma^{\lambda/4}_Y$. Similarly, $\lambda$ can be obtained from an amplitude damping experiment: since $\bra{1} \Lambda^\lambda_\mathrm{damp}(\ket{1}\bra{1})\ket{1} = e^{-\lambda}$ we straightforwardly have $\lambda = t/T_1$.

If we have both dephasing and amplitude damping noise, we can combine the two together as follows. For some $T_1,T_2$, consider the combined noise channel $\Lambda^t_\mathrm{noise} = \exp\left( \frac{t}{T_1} \mathcal{L}_\mathrm{damp} + \frac{t}{T_2}\mathcal{L}_\mathrm{ph}  \right)$. Then:
    \begin{align}
        \tilde{\Lambda}^t_\mathrm{noise} = \Gamma_{X}^{\frac{t}{4T_1}}\Gamma_{Y}^{\frac{t}{4T_1}}\Gamma_{Z}^{\frac{t}{2T_2}}\;.
    \end{align}
This follows from the fact that $\mathcal{L}_\mathrm{damp}$ and $\mathcal{L}_\mathrm{ph}$ commute.

\paragraph*{Noise from unitary gates.} In principle we could perform experiments, as in \cite{Berg2022}, to determine the exact Pauli rates for each unitary, as is necessary for probabilistic error cancellation. However, two-qubit gates like the CNOT gate have fifteen noise parameters corresponding to the $4^2-1$ nontrivial two-qubit Pauli operators. For our purposes we would prefer to model CNOT noise using just a single number. 

One approach could be to just assume that the CNOT noise is simply depolarizing noise. In this case, all fifteen Pauli noise rates are equal and can be connected to the process fidelity. Say we aim to implement an ideal unitary $U$, but our hardware can only implement $\bar{\mathcal{U}} = \mathcal{U}\Lambda_{\mathrm{dep},q}$ up to a known fidelity $F(\mathcal{U},\bar{\mathcal{U}})$. Then  $q = (4^nF(\mathcal{U},\bar{\mathcal{U}}) - 1)/(4^n - 1).$

However, it turns out that spreading out the error uniformly over all the Paulis is rather cumbersome because it requires propagating every possible Pauli error. A more tractable approach is to just consider the worst case Pauli error. In that case,  For any unitary $U$ and $P \in \mathcal{P}$, we have $F(\mathcal{U}, \mathcal{U}\Gamma_P^\lambda) = (1+e^{-2\lambda})/2$. 

\paragraph*{Conclusions.} We have derived a rigorous justification for a rather simple strategy for deriving theoretical predictions of noisy superconducting quantum hardware. Expressions for noise as a function of circuit size can be derived simply by counting the amount of idle time, CNOT gates, and number of mid-circuit measurements. The model has very few parameters, which are simply the Pauli-Lindblad noise rates corresponding to each of these operations (sometimes per unit time). These different noise rates are added up and converted to a fidelity via Lemma~\ref{lemma:product_fidelity}.

The advantage of a rigorous derivation is that we can directly see the ways in which this model fails to tightly capture the actual error. A central issue is that Lemma~\ref{lemma:product_fidelity} does not take into account cancellation between various noise sources, causing the fidelity to approach zero in the limit of high rate. This is despite the fact that the worst possible process fidelity is nonzero. Another oversimplification is that we do not capture the fact that not all possible Pauli noise rates can affect a given observable. We also cannot capture correlations between errors, as may be the case with crosstalk, and instead take a worst-case approach reminiscent of the union-bound. All of these reasons indicate that this model should produce relatively loose lower bounds.
\begin{proof}[\indent Proof of Lemma~\ref{lemma:product_fidelity}.] Say $\Lambda(\rho) = \sum_{P,Q} c_{P,Q} P\rho Q$. Then $\mathcal{F}_\mathrm{proc}(I, \Lambda) = \mathcal{F}_\mathrm{proc}(I, \tilde\Lambda) =  c_{I,I}$.

The proof proceeds with two loose lower bounds that notably fail to capture cancellations between different error sources. 
 Given $\Lambda = \prod_{P} \Gamma_P^{\lambda_P}$, recall that $\Gamma_P^{\lambda_P}(\rho) = (1-\omega_P)\rho + \omega_P P\rho P^\dagger$. Expanding out $\Lambda$, we see that:
    \begin{align}
        c_{I,I} \geq \prod_{P} (1-\omega_P) = \prod_{P} \frac{1+e^{-2\lambda_P}}{2}\;. \label{eqn:cIIexpansion}
    \end{align}
    Next, observing that $(1+e^{-2x})/2 \geq e^{-x}$ for $x > 0$: 
    \begin{align}
        ... \geq \prod_{P} e^{-\lambda_P} = \exp\left(-\sum_P \lambda_P\right)\;.
    \end{align}
\end{proof}

\paragraph*{Convergence to 0.4.} In the main text, we remarked that the fidelities of the measurement-based CNOT experiments converge to a value slightly below 0.4, as is observed in Figure~\ref{fig:cnot} (c). As discussed, this is due to the structure of the measurement-based circuit in Figure~\ref{fig:cnot} (a). While the circuit also experiences infidelity on the top and bottom qubits due to idle time and some CNOTs, the only infidelity that actually scales with $n$ is due to incorrect $Z$ and $X$ corrections on the top and bottom qubits respectively.

We can model this noise as $\Gamma_{ZI}^{\lambda_{ZI}}\Gamma_{IX}^{\lambda_{IX}}$ in the limit of large $\lambda_{ZI},\lambda_{IX}$, in which case $\omega_{ZI},\omega_{IX}$ approach $1/2$. We proceed as in (\ref{eqn:cIIexpansion}). Since these Pauli errors cannot cancel, the calculation is exact.
  \begin{align}
        \mathcal{F}_\mathrm{proc}(I,\Gamma_{ZI}^{\lambda_{ZI}}\Gamma_{IX}^{\lambda_{IX}}) = c_{I,I} = (1-\omega_{ZI})(1 - \omega_{IX}) = 1/4\;.
    \end{align}
This converts to $\mathcal{F}_\mathrm{gate}(I,\Gamma_{ZI}^{\lambda_{ZI}}\Gamma_{IX}^{\lambda_{IX}}) = (4\mathcal{F}_\mathrm{proc}(I,\Gamma_{ZI}^{\lambda_{ZI}}\Gamma_{IX}^{\lambda_{IX}}) + 1)/(4+1) = 0.4$.

\end{document}